\documentclass[11pt,a4paper]{article}
\usepackage{jheppub}
\pdfoutput=1
\usepackage{subcaption}
\usepackage[T1]{fontenc}
\usepackage[utf8]{inputenc}
\usepackage{braket}
\usepackage{slashed}
\usepackage{placeins}
\usepackage{xcolor}
\usepackage{multirow}

\DeclareRobustCommand{\Eq}[1]{Eq.~\eqref{eq:#1}}
\DeclareRobustCommand{\eq}[1]{eq.~\eqref{eq:#1}}
\DeclareRobustCommand{\Eqs}[2]{Eqs.~\eqref{eq:#1} and \eqref{eq:#2}}
\DeclareRobustCommand{\eqs}[2]{eqs.~\eqref{eq:#1} and \eqref{eq:#2}}

\DeclareRobustCommand{\fig}[1]{Fig.~\ref{fig:#1}}
\DeclareRobustCommand{\figs}[2]{Figs.~\ref{fig:#1} and \ref{fig:#2}}
\DeclareRobustCommand{\app}[1]{appendix~\ref{app:#1}}
\DeclareRobustCommand{\sec}[1]{Sec.~\ref{sec:#1}}

\DeclareRobustCommand{\tbl}[1]{Table~\ref{tbl:#1}}
\DeclareRobustCommand{\refcite}[1]{Ref.~\cite{#1}}
\DeclareRobustCommand{\refcites}[1]{Refs.~\cite{#1}}

\newcommand{\MS}{{\overline{\mathrm{MS}}}}
\newcommand{\df}{\mathrm{d}}
\newcommand{\img}{\mathrm{i}}
\newcommand{\eps}{\epsilon}
\newcommand{\cA}{\mathcal{A}}

\newcommand{\cI}{\mathcal{I}}
\newcommand{\cL}{\mathcal{L}}
\newcommand{\cO}{\mathcal{O}}

\newcommand{\nn}{\nonumber}
\newcommand{\bn}{{\bar{n}}}
\newcommand{\as}{\alpha_s}

\newcommand{\Tr}{\mathrm{Tr}}

\newcommand{\bt}{\vec b_T}
\newcommand{\kt}{\vec k_T}

\newcommand{\qt}{\vec q_T}
\newcommand{\ns}{{\rm ns}}
\newcommand{\pb}{{p{\cdot}b}}
\newcommand{\cR}{{\tilde R}}

\title{One-loop Matching for Spin-Dependent Quasi-TMDs\!\!\!\!}

\author[a]{Markus A.~Ebert,}
\author[a]{Stella T.~Schindler,}
\author[a]{Iain W.~Stewart,}
\author[a,b]{and Yong Zhao}

\affiliation[a]{Center for Theoretical Physics, Massachusetts Institute of Technology, Cambridge, MA 02139, USA}
\affiliation[b]{Physics Department, Brookhaven National Laboratory, Bldg. 510A, Upton, NY 11973, USA}%

\emailAdd{ebert@mit.edu}
\emailAdd{stellas@mit.edu}
\emailAdd{iains@mit.edu}
\emailAdd{yzhao@bnl.gov}

\abstract{
Transverse momentum dependent parton distribution functions (TMDPDFs) provide a unique probe of the three-dimensional spin structure of hadrons.
We construct spin-dependent quasi-TMDPDFs that are amenable to lattice QCD calculations and that can be used to determine spin-dependent TMDPDFs. 
We calculate the short-distance coefficients connecting spin-dependent TMDPDFs and quasi-TMDPDFs at one-loop order.
We find that the helicity and transversity distributions have the same coefficient as the unpolarized TMDPDF.  
We also argue that the same is true for pretzelosity and that this spin universality of the matching will hold to all orders in $\alpha_s$.
Thus, it is possible to calculate ratios of these distributions as a function of longitudinal momentum and transverse position utilizing simpler Wilson line paths than have previously been considered.  
}

\preprint{\vbox{
\hbox{MIT--CTP 5200}
}}

\arxivnumber{}

\begin{document}

\maketitle

\section{Introduction}

Understanding the internal structure of hadrons has been a decades-long quest in nuclear and particle physics. Tremendous progress has been made towards the measurement of momentum distributions of quarks and gluons, or parton distribution functions (PDFs), in the longitudinal direction. Multiple experiments that have been carried out recently or that are coming online soon 
promise to open up a new window into the full three-dimensional internal dynamics of nucleons~\cite{Gautheron:2010wva,Dudek:2012vr,Aschenauer:2015eha,Accardi:2012qut}. 
One key target of these experiments are transverse momentum dependent PDFs (TMDPDFs), which provide information about partons in transverse-momentum space and make accessible new correlations between the partonic and hadronic spin. Although TMDPDFs are harder to measure experimentally than their longitudinal counterparts,
a significant effort has built up in recent years to fit these quantities with global data from Drell-Yan and semi-inclusive deep inelastic scattering processes~\cite{Bacchetta:2017gcc,Scimemi:2017etj,Bertone:2019nxa,Scimemi:2019cmh,Bacchetta:2019sam,1793441}. 
It is crucial to develop a deeper theoretical knowledge of TMDPDFs to complement ongoing experimental and global fit analyses and to enhance our understanding of these fundamental hadron properties.

A challenging regime for theoretical calculation of TMDPDFs is at small transverse momentum $q_T \sim \Lambda_{\rm QCD}$, where the TMDPDF is intrinsically nonperturbative and can only be determined from first-principles calculations. Lattice gauge theory is the only known systematic approach to calculate nonperturbative QCD matrix elements, motivating the formulation of TMDPDFs in a manner that is tractable for a Euclidean lattice.
The time-dependence of the nonlocal Wilson-line operators that define TMDPDFs makes this task tricky. The large-momentum effective theory (LaMET) has been proposed  as a method to circumvent this problem by calculating general partonic structures from boosted hadron matrix elements in lattice QCD~\cite{Ji:2013dva,Ji:2014gla,Ji:2020ect}.

This idea motivated investigations into the application of LaMET for TMDPDFs~\cite{Ji:2014hxa,Ji:2018hvs,Ebert:2018gzl,Ebert:2019okf,Ebert:2019tvc,Ji:2019sxk,Ji:2019ewn,Vladimirov:2020ofp,Ji:2020ect}.
Within the formulation of LaMET, one constructs so-called quasi-TMDPDFs that are computable on the lattice and that are related to the desired TMDPDFs through a factorization formula. 
Currently this factorization formula requires the lattice calculation of an additional nonperturbative factor $g_S$~\cite{Ebert:2019okf} or ``reduced soft function''~\cite{Ji:2019sxk}, both of which pose significant challenges and are still not available in the literature.
This complexity can be avoided by forming ratios of quasi-TMDPDFs in different hadron states or for different spin-dependent structures~\cite{Ebert:2019okf}, an idea that has already been utilized for a number of years to access ratios of the $x$-moments of TMDPDFs from similar matrix elements~\cite{Musch:2010ka,Musch:2011er,Engelhardt:2015xja,Yoon:2016dyh,Yoon:2017qzo}. 
In particular, a method was recently constructed to determine the nonperturbative Collins-Soper kernel that governs the energy evolution of TMDPDFs from the ratio of quasi-TMDPDFs at two different hadron momenta~\cite{Ebert:2018gzl,Ebert:2019tvc}. A full practical implementation of this proposal has been realized in exploratory lattice calculations~\cite{Shanahan:2019zcq,Shanahan:2020zxr}. 
A key ingredient in the factorization formula that relates quasi-TMDPDFs and TMDPDFs is the perturbative short-distance matching kernel, which Refs.~\cite{Ji:2018hvs,Ebert:2019tvc} calculate to one-loop order for the unpolarized non-singlet case. 

This paper generalizes the quasi-TMDPDF analysis to the full set of leading-power spin structures and determines the corresponding matching kernels at one-loop order.
Our results enable TMDPDF lattice calculations for the ratios of spin-dependent and unpolarized TMDPDFs. 
Importantly, we find that the matching kernel for TMDPDFs at leading power is the same for all spin structures up to ${\cal O}(\alpha_s)$, and we provide additional arguments that this will remain true to all orders in perturbation theory. 
We also compare two definitions of the quasi-TMDPDF, namely those in~\refcites{Ji:2018hvs,Ebert:2019okf} and in~\refcites{Ji:2019ewn,Ji:2020ect}, which agree in the infinite staple length limit, but differ by where the quark fields are located. 

This paper opens with an introduction to spin-dependent quasi-TMDPDFs in \sec{qTMDdef}. In \sec{oneloop} we calculate the spin-dependent quasi-TMDPDFs at one-loop, comparing them with the spin-dependent TMDPDFs to obtain the matching coefficients. Next, we  discuss potential applications of these calculations in \sec{applications}. We conclude in \sec{conclusion}. Appendix~\ref{app:sym} and \sec{qTMDdef} discuss the advantages of the alternative definition of the quasi-TMDPDF.  In \app{SDTMDPDF} we further detail the one-loop calculation of the spin-dependent quasi-TMDPDFs.

\section{Definition of spin-dependent TMDPDFs and quasi-TMDPDFs}
\label{sec:qTMDdef}

We begin by reviewing the definitions of the spin-dependent TMDPDFs
and their corresponding spin-dependent quasi-TMDPDFs, following the notation of
\refcite{Ebert:2019okf}. We focus on the TMDPDF for an energetic hadron $h$
moving along the $n$-direction, choosing lightlike reference vectors 
\begin{align}
 n^\mu = (1,0,0,1) \,,\qquad \bn^\mu = (1,0,0,-1) \,,\qquad n \cdot \bn = 2
\,,\end{align}
such that the hadron momentum is $P^\mu = \frac12 (P^- n^\mu + P^+ \bn^\mu)$
with $P^- \gg P^+ =M^2/P^-$, where $M$ is the hadron mass. 

Our convention for the lightcone decomposition of momenta is%
\footnote{Another popular convention in the literature is to choose $n_+^\mu = n^\mu/\sqrt{2}$ and $n_-^\mu = \bn^\mu/\sqrt{2}$, such that the lightcone decomposition reads $k^\mu = k^+ n_+^\mu + k^- n_-^\mu + k_\perp^\mu$, where now $k^\pm = (k^0 \pm k^z)/\sqrt{2}$.}
\begin{align}
 k^\mu = k^- \frac{n^\mu}{2} + k^+ \frac{\bn^\mu}{2} + k_\perp^\mu
\,,\end{align}
where $k^+ = n \cdot k = k^0 - k^z$ and  $k^- = \bn \cdot k = k^0 + k^z$. The transverse component of the four-vector $k^\mu$ is
$k_\perp^\mu = (0, \kt, 0)$ with $k_\perp^2 = -\kt^2 \equiv -k_T^2$.
It is also useful to define the transverse metric $g_\perp^{\mu\nu}$
and the transverse antisymmetric tensor $\eps^{\mu\nu} \equiv \eps_\perp^{\mu\nu}$ as
\begin{align}
 g_\perp^{\mu\nu} = g^{\mu\nu} - \frac12 \bigl( n^\mu \bn^\nu + \bn^\mu n^\nu\bigr)
\,,\qquad
 \eps^{\mu\nu} = \frac12 n_\alpha \bn_\beta \eps^{\alpha \beta \mu\nu}
\,,\end{align}
such that $g_\perp^{11} = g_\perp^{22} = -1$ and $\eps^{12} = - \eps^{21} = 1$.

\subsection{Definition of spin-dependent TMDPDFs}
\label{sec:def_TMDPDF}

The TMDPDF for finding a quark of flavor $q$ inside a hadron $h$ with momentum $P$ and polarization $S$ is defined as
\begin{align} \label{eq:tmdpdf}
 f_{q/h_S}^{[\Gamma]}(x, \bt, \mu, \zeta) &
 = \lim_{\substack{\eps\to0 \\ \tau\to0}} Z^q_{\rm uv}(\mu,\zeta,\eps) \,
   B_{q/h_S}^{[\Gamma]}\bigl(x, \bt, \eps, \tau, xP^- \bigr)
   \frac{\sqrt{S^q(b_T,\eps,\tau)}}{S_q^0(b_T,\eps,\tau)}
\,,\nn\\ &
 \equiv \lim_{\substack{\eps\to0 \\ \tau\to0}} Z^q_{\rm uv}(\mu,\zeta,\eps) \,
   B_{q/h_S}^{[\Gamma]}\bigl(x, \bt, \eps, \tau, xP^- \bigr)
   \Delta_S^q(b_T,\eps,\tau)
\,.\end{align}
Here, $x$ is the fraction of the proton momentum $P^-$ carried by the struck quark,
$\bt$ is Fourier-conjugate to the quark's transverse momentum, and $\mu$ and $\zeta$
are the renormalization and Collins-Soper scale $\zeta$~\cite{Collins:1981va,Collins:1981uk}, respectively.
The Dirac structure $\Gamma$ projects out the desired spin-dependent TMD, as discussed below.
In \eq{tmdpdf}, $Z^q_{\rm uv}$ is the UV counterterm in the $\MS$~scheme, where UV divergences are regulated by working in $d=4-2\eps$ dimensions.
The beam function or unsubtracted TMDPDF $B_{q/h_S}^{[\Gamma]}$
is a hadronic matrix element. The soft function $S^q$ is a vacuum matrix element and thus is independent of the hadron state, hadron momentum and spin, and quark flavor $q$ and spin being probed.
$S_q^0$ subtracts the overlap between the soft and beam functions (and is sometimes referred to as a zero-bin subtraction~\cite{Manohar:2006nz}). Its precise definition depends on the choice of a rapidity regulator $\tau$.
The second line of \eq{tmdpdf} incorporates the soft function and the zero-bin into a combined soft factor $\Delta_S^q$.

A characteristic feature of TMDPDFs is the appearance of so-called rapidity divergences in the beam and soft functions,
such that both need an additional rapidity regulator generically denoted as $\tau$ in \eq{tmdpdf}.
The divergences cancel in the combination in \eq{tmdpdf}, such that the regulator can be removed as $\tau\to0$, yielding a regulator-independent definition of the TMDPDF.
Analogous to the appearance of the renormalization scale $\mu$, the rapidity divergences induce the Collins-Soper scale $\zeta$, which is related to the momentum of the struck quark by $\zeta \propto (x P^-)^2$.
The proportionality constants depends on the chosen rapidity regulator $\tau$, and many different schemes are known in the literature, such as taking Wilson lines off the light cone~\cite{Collins:1350496}, analytic regulators~\cite{Beneke:2003pa, Chiu:2007yn, Becher:2011dz}, the $\eta$-regulator~\cite{Chiu:2011qc, Chiu:2012ir}, the $\delta$ regulator \cite{Chiu:2009yx, GarciaEchevarria:2011rb}, and the exponential regulator~\cite{Li:2016axz}.
(We do not consider the regulators of \refcites{Collins:1981uk,Ji:2004wu}, which induce an additional regularization parameter $\rho$; using these to calculate cross-sections gives rise to the product of a TMDPDF and a hard coefficient that are not in the $\MS$~scheme.)

\begin{figure*}
 \centering
 \includegraphics[width=0.4\textwidth]{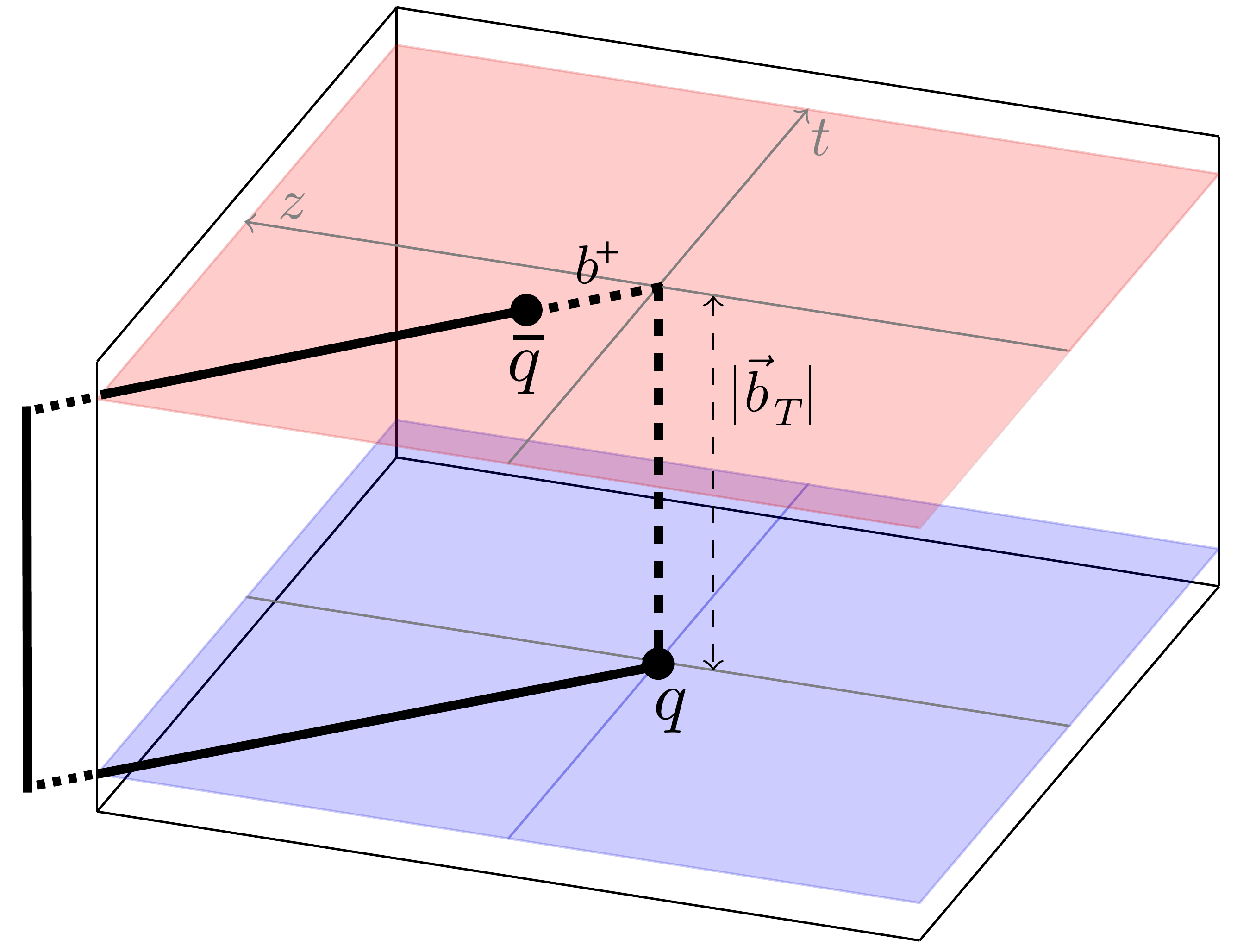}
 \qquad
 \includegraphics[width=0.4\textwidth]{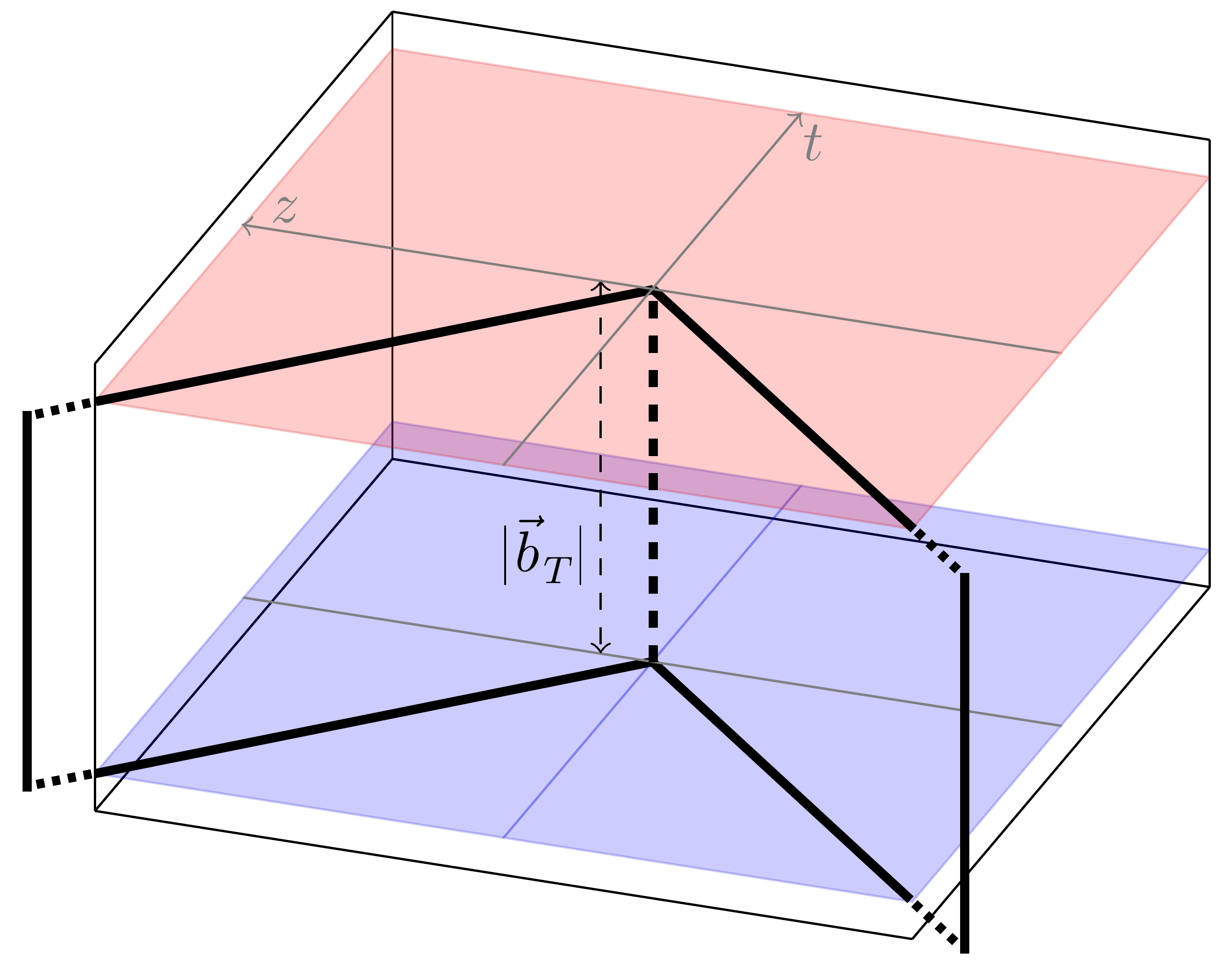}
 \caption{Graphs of the Wilson line structure of the $n$-collinear beam function $B_{q}$ (left)
 and the soft function $S^q$ (right), defined in \eqs{beamfunc}{softfunc}.
 The Wilson lines (solid) extend to infinity in the directions indicated.
 Adapted from \refcite{Li:2016axz}.}
 \label{fig:wilsonlines}
\end{figure*}

We define the bare beam and soft functions as
\begin{align} \label{eq:beamfunc}
 B_{q/h_S}^{[\Gamma]}(x,\bt,\eps,\tau,x P^-) &
 = \int\frac{\df b^+}{4\pi} e^{-\img \frac12 b^+ (x P^-)}
 \Bigl< h_S(P) \Bigr|  \Bigl[ \bar q(b^\mu) W_{\sqsubset}(b^\mu,0) \frac{\Gamma}{2} 
  q(0) \Bigr]_{\tau,\epsilon} \Bigl| h_S(P) \Bigr>
\,,\\   \label{eq:softfunc} 
 S^q(b_T,\eps,\tau) &= \frac{1}{N_c} \bigl< 0 \bigr| \Tr \bigl[ S^\dagger_n(\bt) S_\bn(\bt)
   S_{T}(-\infty \bn;\vec b_T,\vec 0_T)
 \nn\\&\hspace{2cm}\times
 S^\dagger_\bn(\vec 0_T) S_n(\vec 0_T)
 S_{T}^\dagger\bigl(-\infty n;\vec b_T,\vec 0_T\bigr) \bigr]_{\tau,\epsilon}
 \bigl|0 \bigr>
\,.\end{align}
Here, we use $[\cdots]_{\tau,\epsilon}$ to denote that the operator in brackets is rapidity-regulated by $\tau$ and that the invariant mass is regulated by dimensional regularization. In \eqs{beamfunc}{softfunc}, we define the Wilson lines  as
\begin{align} \label{eq:wilsonlines}
  W_{\sqsubset}(b^\mu,0) &=
 W(b^\mu) W_{T}\bigl(-\infty\bn;\vec b_T,\vec 0_T\bigr)W^\dagger(0)
\,,\nn\\
 W(x^\mu) &
 = P \exp\biggl[ -\img g \int_{-\infty}^0 \df s\, \bn \cdot \cA(x^\mu + s \bn^\mu) \biggr]
\,,\nn\\
 S_n(x^\mu) &
 = P \exp\biggl[ -\img g \int_{-\infty}^0 \df s\, n \cdot \cA(x^\mu + s n^\mu) \biggr]
\,,\nn\\
 W_{T}(x^\mu;\vec b_T,\vec 0_T) &
 = P \exp\left[ \img g \int_{\vec 0_T}^{\vec b_T} \df \vec s_T \cdot \vec \cA_T(x^\mu + s_T^\mu) \right]
 = S_{T}(x^\mu;\vec b_T,\vec 0_T)
\,.\end{align}
Note that the transverse gauge links at light-cone infinity are required to obtain the connected Wilson line paths necessary for gauge-invariant matrix elements.
We can often neglect the transverse gauge links in nonsingular gauges such as Feynman gauge, where the gluon field strength vanishes at infinity. 
However, they are important in certain singular gauges, see e.g.\ \refcites{Ji:2002aa,Belitsky:2002sm,Idilbi:2010im,GarciaEchevarria:2011md}, and for operators that are formulated for a finite-size lattice.
We illustrate the Wilson paths of the matrix elements in \eqs{beamfunc}{softfunc} in \fig{wilsonlines}.

At leading power (often referred to as twist-2), only three Dirac structures contribute to the TMDPDF:
\begin{align} \label{eq:Gamma_LP}
 \Gamma \in \{\, \slashed{\bn} \,,\, \slashed\bn\gamma_5 \,,\, \img \sigma^{\alpha-}\gamma_5 \,\}
\,,\end{align}
where $\sigma^{\mu\nu} = \frac{\img}{2} [\gamma^\mu, \gamma^\nu]$.
The resulting spin-dependent TMDs for a spin-$\frac12$ hadron comprise eight independent Lorentz structures~\cite{Ralston:1979ys,Tangerman:1994eh,Boer:1997nt,Mulders:1995dh}
(see also \refcites{Bacchetta:2004zf,Goeke:2005hb,Bacchetta:2006tn} for a complete classification at higher orders in the power expansion, twist-3 and twist-4).
Following the notation of~\refcite{Bacchetta:2006tn}, the leading power decomposition in momentum space is%
\footnote{Note that these expressions are identical to $\Phi^{[\gamma^+]}, \Phi^{[\gamma^+ \gamma_5]}, \Phi^{[\img \sigma^{\alpha+}\gamma_5]}$ in \refcite{Bacchetta:2006tn}, since the factor $1/\sqrt{2}$ for converting the different lightcone conventions is already accounted for in \eq{beamfunc}.}
\begin{align} \label{eq:tmd_decomposition}
 f_{q/h_S}^{[\slashed\bn]}(x,\qt) &
 = f_1(x,q_T)
 - \frac{\eps_{\rho\sigma} q_\perp^\rho S_\perp^\sigma}{M} f_{1T}^\perp(x,q_T)
\,,\nn\\
 f_{q/h_S}^{[\slashed\bn\gamma_5]}(x,\qt) &
 = S_L \, g_{1L}(x,q_T)
 - \frac{q_\perp \cdot S_\perp }{M} g_{1T}(x,q_T)
\,,\nn\\
 f_{q/h_S}^{[\img \sigma^{\alpha-}\gamma_5]}(x,\qt) &
 = S_\perp^\alpha h_1(x,q_T)
 + \frac{S_L q_\perp^\alpha}{M} h_{1L}^\perp(x,q_T)
 \nn\\&\quad
 + \frac{q_\perp^2}{M^2} \biggl(\frac12 g_\perp^{\alpha\rho} - \frac{q_\perp^\alpha q_\perp^\rho}{q_\perp^2}\biggr) S_{\perp\,\rho} h_{1T}^\perp(x,q_T)
 - \frac{\eps^{\alpha\rho} q_{\perp\rho}}{M} h_1^\perp(x,q_T)
\,.\end{align}
Here, we drop the arguments $\mu$ and $\zeta$ for brevity. We also distinguish the position-space TMD $f_{q/h_S}^{[\Gamma]}(x,b_T)$ and momentum-space TMD $f_{q/h_S}^{[\Gamma]}(x,\qt)$ only by their arguments.
Note that all functions on the right-hand side of \eq{tmd_decomposition} only depend on the magnitude $q_T = |\qt|$ of the transverse momentum.
In \eq{tmd_decomposition}, $M$ denotes the nucleon mass, $\alpha$ is a tranverse index, and the spin vector of the hadron $h$ is decomposed as
\begin{align} \label{eq:spin_vector}
 S^\mu
 = S_L \frac{ P^- n^\mu - P^+ \bn^\mu}{2 M} + S_\perp^\mu
\,,\end{align}
where $-S_L^2 + S_\perp^2 = -1$ such that $S^2 = -1$.
\Eq{tmd_decomposition} contains eight distributions, which each
correspond to a specific choice of polarizations for a quark and its parent hadron, as summarized in \tbl{tmds}.
Six of these distributions are time-reversal even (T-even); namely, the unpolarized ($f_1)$, helicity ($g_{1L}$), transversity ($h_1$), pretzelosity ($h_{1T}^\perp$) and worm-gear ($g_{1T}$ and $h_{1L}^\perp$) distributions.
The other two are T-odd; specifically, the Sivers ($f_{1T}^\perp$)~\cite{Sivers:1989cc} and Boer-Mulders ($h_1^\perp$)~\cite{Boer:1997nt} functions.

\renewcommand{\arraystretch}{1.2}
\begin{table}[t]
 \centering
 \begin{tabular}{|*{5}{c|}}
 \cline{3-5}
 \multicolumn{2}{c|}{} & \multicolumn{3}{c|}{\textbf{Quark polarization}}
 \\ \cline{3-5}
 \multicolumn{2}{c|}{} & \hspace{0.5cm}U\hspace{0.5cm} & \hspace{0.5cm}L\hspace{0.5cm} & \hspace{0.5cm}T\hspace{0.5cm}
 \\ \hline
 \multirow{3}{*}{\parbox{1.5ex}{\vspace{0.15cm}\rotatebox[origin=c]{90}{\textbf{Hadron polarization}}}}
 & \parbox{1.5ex}{\vspace{3ex}U} & $f_1$ & & $h_1^\perp$
 \\ & & \small unpolarized & & \small Boer-Mulders
 \\ \cline{2-5}
 & \parbox{1.5ex}{\vspace{3ex}L} & & $g_{1L}$ & $h_{1L}^\perp$
 \\ & &  & \small helicity & \small worm-gear
 \\ \cline{2-5}
 & \parbox{1.5ex}{\vspace{3ex}T} & $f_{1T}^\perp$ & $g_{1T}$ & $h_{1}$, $h_{1T}^\perp$
 \\ & & \small Sivers & \small worm-gear & \small transversity, pretzelosity
 \\ \hline
 \end{tabular}
 \caption{Overview of the quark TMDs of a spin-$\frac12$ hadron at leading power in \eq{tmd_decomposition}, organized by the polarization of the struck quark and the parent hadron. Here U, L, T stand for unpolarized, longitudinally polarized and transversely polarized, respectively. The Boer-Mulders and Sivers functions are time-reversal odd, while all other functions are time-reversal even.
 }
 \label{tbl:tmds}
\end{table}
\renewcommand{\arraystretch}{1.0}

One can obtain the position-space version of \eq{tmd_decomposition} by a Fourier transform with respect to $\qt$.
As in \refcite{Gutierrez-Reyes:2017glx}, we normalize the position-space functions so that explicit factors of $M$ are absent.
With this choice, we have
\begin{align} \label{eq:tmd_decomposition_2}
 f_{q/h_S}^{[\slashed\bn]}(x,\bt) &
 = f_1(x,b_T)
 - \frac{\eps_{\rho\sigma} b_\perp^\rho S_\perp^\rho}{b_T} f_{1T}^\perp(x,b_T)
\,,\nn\\
 f_{q/h_S}^{[\slashed\bn\gamma_5]}(x,\bt) &
 = S_L g_{1L}(x,b_T)
 - \frac{b_\perp \cdot S_\perp}{b_T} g_{1T}(x,b_T)
\,,\nn\\
 f_{q/h_S}^{[\img \bn_\beta \sigma^{\alpha\beta}\gamma_5]}(x,\bt) &
 = S_\perp^\alpha h_1(x,b_T)
 + \frac{S_L b_\perp^\alpha}{b_T} h_{1L}^\perp(x,b_T)
 \nn\\&\quad
 + \biggl(\frac12 g_\perp^{\alpha\rho} - \frac{b_\perp^\alpha b_\perp^\rho}{b_\perp^2}\biggr) S_{\perp\,\rho} h_{1T}^\perp(x,b_T)
 - \frac{\eps^{\alpha\rho} b_{\perp\rho}}{b_T} h_1^\perp(x,b_T)
\,,\end{align}
where $b_\perp^\mu = (0,\bt,0)$ and $b_T = |\bt|$.
The explicit relations between position-space and momentum-space distributions are given by (see also \refcite{Echevarria:2015uaa})
\begin{align} \label{eq:ft}
 F(x,b_T) &= 2\pi \int_0^\infty \df k_T \, k_T J_0(b_T k_T) F(x, k_T)
 \,,\hspace{2cm} F \in \{ f_1, g_{1L}, h_1 \}
\nn\\
 F(x,b_T) &= 2\pi\img \int_0^\infty \df k_T \, k_T \frac{k_T}{M} J_1(b_T k_T) F(x,k_T)
 \,,\hspace{1.36cm} F \in \{ f_{1T}^\perp, g_{1T}, h_{1L}^\perp, h_1^\perp \}
\nn\\
 h_{1T}^\perp(x,b_T) &= -2\pi \int_0^\infty \df k_T \, k_T \frac{k_T^2}{M^2} J_2(b_T k_T) h_{1T}^\perp(x,k_T)
\,,\end{align}
where $J_n(x)$ is a Bessel function of the first kind.
Note that only $f_1$, $g_{1L}$ and $h_1$ are directly related by a Fourier transform to their momentum-space counterparts. The remaining functions require higher-order Bessel transforms reweighted with an appropriate factor of $(k_T/M)$ or $(k_T/M)^2$, due to the normalization conventions in \eq{tmd_decomposition}.

\subsection{Definition of spin-dependent quasi-TMDPDFs}
\label{sec:def_qTMDPDF}

Within the framework of LaMET~\cite{Ji:2013dva,Ji:2014gla}, quasi-TMDPDFs are defined as equal-time analogs of TMDPDFs, and are thus amenable to lattice calculations.
To define the spin-dependent quasi-TMDPDFs, we extend the setup in ~\refcite{Ebert:2019okf}:
\begin{align} \label{eq:qtmdpdf}
 \tilde f_{q/h_S}^{[\tilde\Gamma]}(x, \bt,\mu,P^z)
 = \int \frac{\df b^z}{2\pi} \, e^{\img b^z (x P^z)}\,
 &\tilde Z'_q(b^z,\mu,\tilde \mu) \tilde Z_{\rm uv}^q(b^z,\tilde \mu, a)
 \nn\\&\times
 \tilde B_{q/h_S}^{[\tilde\Gamma]}(b^z, \bt, a, L, P^z) \tilde\Delta_S^q(b_T, a, L)
\,.\end{align}
Here, $h_S$ denotes a hadron with spin $S$ and momentum $P$, $q$ is the flavor of the struck quark, $x$ is the fraction of the $z$-momentum carried by this quark, and $\bt$ is Fourier-conjugate to the quark's transverse momentum. The scale $\mu$ is the $\MS$-renormalization scale, and the $P^z$ dependence here plays the role of the Collins-Soper scale $\zeta$.
Moreover, $\tilde B_{q/h_S}^{[\tilde\Gamma]}$ and $\tilde\Delta_S^q$ denote the quasi beam function and quasi soft factor. 

For the purposes of this paper, we need not go into details about the precise definition of $\tilde\Delta_S^q$; while we must include it in the
quasi beam function to cancel $L/b_T$ divergences, it cancels out when taking the ratio of two quasi-TMDPDFs.
In \eq{qtmdpdf} the letter $a$ denotes the UV regulator, resembling the notation for the lattice spacing that acts as a UV regulator in practical calculations. Here the rapidity regulator is replaced by the dependence on the length $L$ of the staple-shaped Wilson lines defined below~\cite{Ji:2018hvs,Ebert:2019okf}. 
In contrast to \eq{tmdpdf}, we formulate the UV renormalization in \eq{qtmdpdf} in $b^z$ space, where the lattice renormalization factor $\tilde Z_{\rm uv}^q$ absorbs Wilson-line self energies proportional to $b^z$, with $\tilde\mu$ being the corresponding renormalization scale.
\footnote{More precisely, $\tilde\mu$ stands for all parameters induced by the lattice renormalization; see e.g.~\refcite{Ebert:2019tvc} for details in the RI$^\prime$/MOM scheme.}
Finally, $\tilde Z_q^\prime$ converts this lattice-renormalization scheme into the $\MS$ scheme.
For more details, we refer to~\refcites{Ebert:2019okf,Ebert:2019tvc}.

\begin{figure*}
 \centering
 \includegraphics[width=0.4\textwidth]{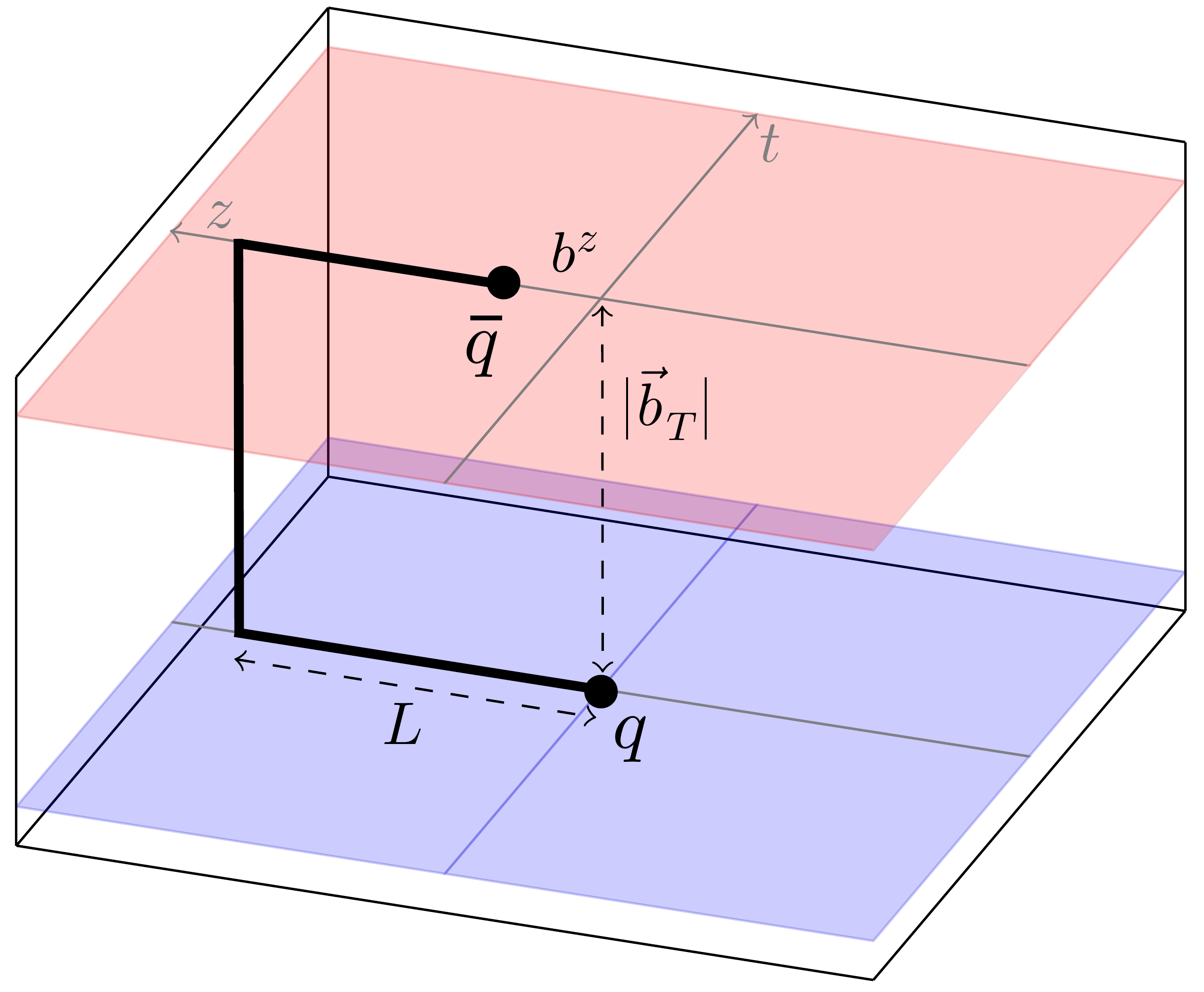}
 \qquad
 \includegraphics[width=0.4\textwidth]{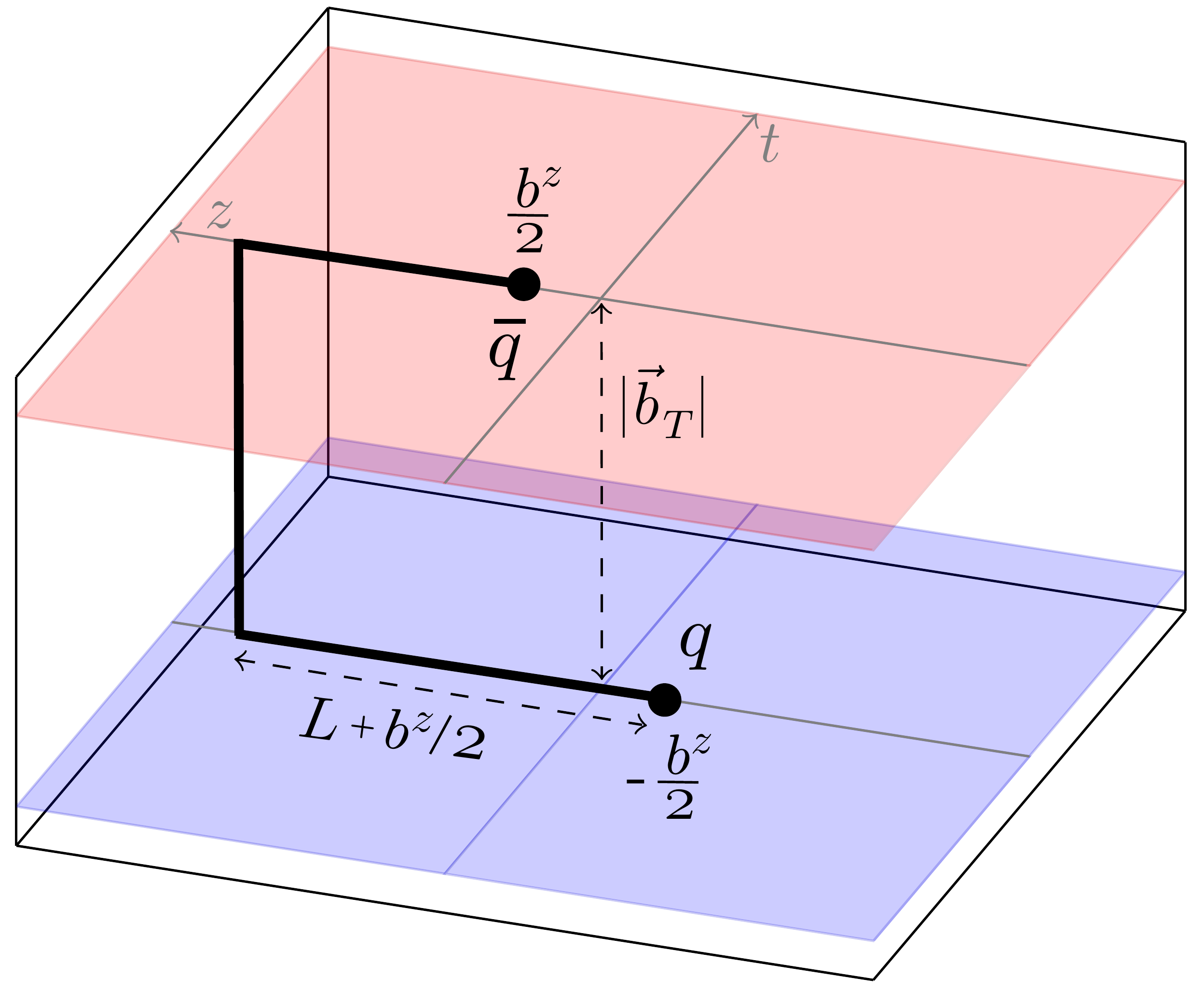}
 \caption{Illustration of the Wilson line structure of the quasi beam function $\tilde B_{q}$ as defined in \eq{qbeam_a} (left) and \eq{qbeam_b} (right).
 }
 \label{fig:qwilsonlines}
\end{figure*}

We define the quasi beam function in position space as
\begin{align} \label{eq:qbeam}
 \tilde B_{q/h_S}^{[\tilde\Gamma]}(b^z, \bt, a,L,P^z) &
 = N_{\tilde\Gamma} \, \Bigl< h_S(P) \Big|\bar q(b^\mu) \widetilde W_{\sqsubset}(b^\mu,0; L) \frac{\tilde\Gamma}{2} q(0) \Big| h_S(P) \Bigr>
\,,\end{align}
where $h_S(P)$ denotes a hadron $h$ with polarization vector $S$ and momentum $P$, and $b^\mu = (0, \bt, b^z)$. Here we follow  \refcite{Ebert:2019okf}. $\widetilde W_{\sqsubset}(b^\mu,0;L)$ is a staple-shaped Wilson line path of length $L$ connecting the quark fields, as illustrated in the left panel of \fig{qwilsonlines}; it is given by
\begin{align} \label{eq:qbeam_a}
  \widetilde W_{\sqsubset}(b^\mu,0; L) &= W_{\hat z}(b^\mu;L-b^z) 
   W_T(L \hat z; \bt, \vec{0}_T) W^\dagger_{\hat z}(0;L) 
\,,\end{align}
where $W_T$ is given in \eq{wilsonlines} and  $W_{\hat z}$ denotes a Wilson line oriented along the $z$-direction,
\begin{align} \label{eq:Wilson_L}
 W_{\hat z}(x^\mu;L) &= P \exp\left[ \img g \int_{L}^0 \df s \, \cA^z(x^\mu + s \hat z) \right]
\,.\end{align}
The Dirac structures $\tilde\Gamma$ come from the set
\begin{align} \label{eq:qGamma}
 \tilde\Gamma \in \{\, \gamma^\lambda \,,\, \gamma^\lambda \gamma_5 \,,\, \img \sigma^{\alpha \lambda} \gamma_5 \,\} \,,\qquad \lambda  = 0 , 3
\,,\end{align}
because they can be boosted to the three corresponding Dirac structures in \eq{Gamma_LP}.
The normalization factor $N_{\tilde\Gamma}$ in \eq{qbeam} is defined as
\begin{align} \label{eq:NGamma}
 N_{\gamma^3} = N_{\gamma^0\gamma_5} = N_{i\sigma^{i3}\gamma_5} = 1
\,,\qquad
 N_{\gamma^0} = N_{\gamma^3\gamma_5} = N_{i\sigma^{i0}\gamma_5} = \frac{P^z}{P^0}
\,.\end{align}
While the choices $\lambda=0$ and $\lambda=3$ are formally equivalent in a continuum analysis,
they induce different operator mixings on a discretized lattice with broken chiral symmetry~\cite{Constantinou:2019vyb,Shanahan:2019zcq,Green:2020xco}. Hence, we consider both options.

Since all choices in \eq{qGamma} boost onto the corresponding Dirac structures in \eq{Gamma_LP}, we can decompose the quasi-TMDs in the same fashion as the TMDs, up to power corrections suppressed by large $P^z$.
Using the conventions of \eq{tmd_decomposition_2}, at leading power we find
\begin{align} \label{eq:qtmd_decomposition}
 \tilde f_{q/h_S}^{[\gamma^\lambda]}(x,\bt) &
 = \tilde f_1^\lambda(x,b_T)
 - \frac{\eps_{\rho\sigma} b_\perp^\rho S_\perp^\sigma}{b_T} \tilde f_{1T}^{\lambda\perp}(x,b_T)
\,,\nn\\
 \tilde f_{q/h_S}^{[\gamma^\lambda\gamma_5]}(x,\bt) &
 = S_L \tilde g_{1L}^\lambda(x,b_T)
 - \frac{b_\perp \cdot S_\perp}{b_T} \tilde g_{1T}^\lambda(x,b_T)
\,,\nn\\
 \tilde f_{q/h_S}^{[\img \sigma^{\alpha\lambda}\gamma_5]}(x,\bt) &
 = S_\perp^\alpha \tilde h_1^\lambda(x,b_T)
 + \frac{S_L b_\perp^\alpha}{b_T} \tilde h_{1L}^{\lambda\perp}(x,b_T)
 \nn\\&\quad
 + \biggl(\frac12 g_\perp^{\alpha\rho} - \frac{b_\perp^\alpha b_\perp^\rho}{b_\perp^2}\biggr) S_{\perp\,\rho} \tilde h_{1T}^{\lambda\perp}(x,b_T)
 - \frac{\eps^{\alpha\rho} b_{\perp\rho}}{b_T} \tilde h_1^{\lambda\perp}(x,b_T)
\,.\end{align}
By choosing the appropriate projector $\tilde\Gamma$ and hadron polarization $S$, it is straightforward to obtain all individual distributions.

Each Wilson line segment gives rise to a power-law divergence proportional to its length due to self-energy contributions~\cite{Dotsenko:1979wb,Craigie:1980qs,Dorn:1986dt}.
As argued in Ref.~\cite{Ebert:2019tvc,Green:2020xco}, the nonlocal Wilson line operator that defines the quasi beam function can be multiplicatively renormalized by the factor
\begin{align}  \label{eq:Zse_a}
	Z(b^z,\tilde \mu,a,b_T,L) = Z_{q,\rm wf}(\tilde \mu,a)\, e^{\delta m (2 L - b^z + b_T)}\,,
\end{align}
where $\delta m=\delta m(a)\sim 1/a$ absorbs the power-law dviergence originating from each Wilson line segment, and $Z_{q,\rm wf}(\tilde \mu,a)$ renormalizes all logarithmic UV divergences from the wave functions and quark-Wilson-line vertices. Thus the $Z$ factor is independent of $\tilde\Gamma$.
The $b^z$ dependence in \eq{Zse_a} induces the presence of $b^z$ in $\tilde Z_{\rm uv}^q(b^z,\tilde \mu,a)$ in \eq{qtmdpdf}, since subtracting the self energies on the lattice necessitates the use of a $b^z$-dependent counterterm. This also leads to a $b^z$-dependent factor $\tilde Z_q'(b^z,\mu,\tilde \mu)$ to convert to the $\MS$ scheme.

To avoid this complication, we can slightly modify the definition in \eq{qbeam_a} so that the length of the Wilson line path is independent of $b^z$.  We can achieve this either by shifting $L \to L + b^z/2$ in \eq{qbeam_a}, or equivalently by using the definition of \refcites{Ji:2019ewn,Ji:2020ect},
\begin{align} \label{eq:qbeam_b}
 \tilde B_{q}^{[\tilde\Gamma]}(b^z, \bt, a,L,P^z) &
 = N_{\tilde\Gamma} \, \Bigl< h_S(P) \Big|\bar q\Bigl(b_\perp + \frac{b^z}{2} \hat z\Bigr) \widetilde W_{\sqsubset}\Bigl(b_\perp+\frac{b^z}{2}\hat z, -\frac{b^z}{2}\hat z; L+\frac{b^z}{2} \Bigr) 
 \,,\nn\\*
 & \qquad\qquad\qquad\ \times \frac{\tilde\Gamma}{2} q\Bigl(-\frac{b^z}{2} \hat z\Bigr) \Big| h_S(P) \Bigr>
\,.\end{align}
We illustrate this modified path in the right panel of \fig{qwilsonlines}.
The operator in \eq{qbeam} keeps the position of one quark field fixed so that only one Wilson line is varied for each choice of $b^z$ on the lattice, whereas all three of the Wilson lines are varied for the operator in \eq{qbeam_b}.

Importantly, \eqs{qbeam_a}{qbeam_b} are equivalent in the limit $L \gg b^z$; thus, all results expanded in this limit apply to both definitions.
The RI/MOM$^\prime$ renormalization factor calculated at one loop in \refcite{Ebert:2019tvc} applies directly to \eq{qbeam_a}, and can be applied to \eq{qbeam_b} after a shift $L \to L + b^z/2$. 

A key benefit of using the definition in \eq{qbeam_b} is that the renormalization factors in \eq{qtmdpdf} become independent of $b^z$ and can be extracted from the Fourier transform, giving us
\begin{align} \label{eq:qtmdpdf_b}
 \tilde f_{q/h_S}^{[\tilde\Gamma]}(x, \bt,\mu,P^z) &
 = \lim_{\substack{a\to0\\L\to\infty}} \tilde Z_q(\mu, a) \tilde B_{q/h_S}^{[\tilde\Gamma]}(x, \bt, a, L, P^z) \tilde\Delta_S^q(b_T, a, L)
\,,\nn\\
 \tilde B_{q/h_S}^{[\tilde\Gamma]}(x, \bt, a, L, P^z) &
 = \int\frac{\df b^z}{2\pi} e^{\img b^z (x P^z)} \tilde B_{q/h_S}^{[\tilde\Gamma]}(b^z, \bt, a, L, P^z)
\,.\end{align}
Here $\tilde Z_q(\mu, a) = \tilde Z'_q(\mu,\tilde \mu) \tilde Z_{\rm uv}^q(\tilde \mu, a)$ is the $b^z$-independent combination of lattice renormalization and $\MS$ conversion factors.
Alternatively, one could of course also apply the renormalization and soft subtraction prior to the Fourier transform.
Moreover, $\tilde Z_q(\mu, a)$ will drop out in the ratios of quasi-TMDPDFs, further simplifying the calculation. 
Appendix~\ref{app:sym} discusses using \eq{qbeam_b} for calculating the Collins-Soper kernel.

\subsection{Relating quasi-TMDPDFs and TMDPDFs}
\label{sec:relation}

Following the notation of \refcite{Ebert:2019okf}, we write the relation between TMDs and quasi-TMDs as
\begin{align} \label{eq:relation}
 \tilde F_{\ns/h_S}(x, b_T, \mu, P^z) &
 = C^{\tilde F}_{\ns}\bigl(\mu, x P^z\bigr) g^S_{q}(b_T, \mu)
   \exp\biggl[ \frac12  \gamma_\zeta^q(\mu, b_T) \ln\frac{(2 x P^z)^2}{\zeta} \biggr]
 F_{\ns/h_S}(x, b_T, \mu, \zeta)
 \nn\\&\quad
 + \cO\biggl(\frac{b_T}{L}, \frac{1}{b_T P^z}, \frac{1}{P^z L}\biggr)
\,.\end{align}
Here, $F$ can be any of the spin-dependent TMDs in \eq{tmd_decomposition_2}, $\tilde F$ is the corresponding quasi-TMD, and the nonsinglet combination $ns=u{-}d$ is chosen to avoid mixing with gluons.
Notably, the relation between TMDs and quasi-TMDs is multiplicative in $x$ space, with $C_{\ns}^{\tilde F}$ being a perturbative kernel. 
It also involves the nonperturbative Collins-Soper evolution to render the right-hand side independent of $\zeta$ for situations where $\zeta\ne (2xP^z)^2$, such as when the quasi-TMD and TMD have different values for $P^z$.
In addition, following \refcite{Ebert:2019okf} we have included in \eq{relation} a nonperturbative factor $g_q^S(b_T,\mu)$ which ensures that this relation is independent of the precise choice for the quasi soft function $\tilde \Delta_S^q(b_T,a,L)$ in \eq{qtmdpdf}.
(See \refcites{Ji:2019ewn,Ji:2020ect,Vladimirov:2020ofp} for discussions of this structure.)
The precise choice for $\tilde\Delta_S^q$, and thus $g_q^S$, is irrelevant for ratios of TMDs and quasi-TMDs, and thus does not impact our results.
\Eq{relation} is correct up to corrections suppressed by large momenta $P^z$ and staple lengths $L$, as indicated.

We consider ratios of \eq{relation} with different TMDs, such that the spin-independent Collins-Soper evolution and $g_q^S$ cancel.
For example, division by the unpolarized TMD gives
\begin{align} \label{eq:relation_ratio}
 \frac{\tilde F_{\ns/h_S}(x, b_T, \mu, P^z)}
      {\tilde f_{\ns}(x, b_T, \mu, P^z)} &
 = \frac{C^{\tilde F}_{\ns}\bigl(\mu, x P^z\bigr)}
        {C_{\ns}\bigl(\mu, x P^z\bigr)}
   \frac{F_{\ns/h_S}(x, b_T, \mu, \zeta)}
        {f(x, b_T, \mu, \zeta)}
\,,\end{align}
up to power-suppressed terms. \refcite{Ebert:2019okf} gives the kernel appearing in the denominator:
\begin{align} \label{eq:kernel_nlo}
 C_{\ns}\bigl(\mu, x P^z\bigr) &
 = 1 + \frac{\as C_F}{4\pi} \biggl[ -\ln^2\frac{(2 x P^z)^2}{\mu^2}
   + 2 \ln\frac{(2 x P^z)^2}{\mu^2} - 4 + \frac{\pi^2}{6} \biggr] + \cO(\as^2)
\,.\end{align}
A main goal of our work here will be to determine results for the kernels $C^{\tilde F}_{\ns}\bigl(\mu, x P^z\bigr)$ appearing in the numerator.

Another advantage of \eq{relation_ratio} is that the soft factor $\tilde\Delta_S^q$ contained in the quasi-TMDs, see \eq{qtmdpdf}, also cancels in the ratio. We may thus equally well write
\begin{align} \label{eq:relation_ratio_2}
 \frac{\tilde B_{\ns/h_S}(x, b_T, \mu, P^z)}
      {\tilde B_{\ns}(x, b_T, \mu, P^z)} &
 = \frac{C^{\tilde F}_{\ns}\bigl(\mu, x P^z\bigr)}
        {C_{\ns}\bigl(\mu, x P^z\bigr)}
   \frac{F_{\ns/h_S}(x, b_T, \mu, \zeta)}
        {f(x, b_T, \mu, \zeta)}
\,,\end{align}
where $\tilde B_{\ns/h_S}$ and $\tilde B_\ns$ are the beam functions that give  rise to the spin-dependent and unpolarized quasi-TMDs $\tilde F_\ns$ and $\tilde f_\ns$, respectively.

\section{One-loop calculation}
\label{sec:oneloop}

With the required formalism in place, we now turn to the calculation of
the matching kernel $C_\ns^{\tilde F}$ appearing in \eq{relation}.
Since the matching results are independent of the precise choice of states, as long as they have overlap with the operators, we can carry out the calculation of the matrix elements in \eqs{beamfunc}{qbeam} using an on-shell quark state $u_s(p)$ with spin vector $s$ and momentum $p^\mu = p^- n^\mu/2$, where $p^- = 2 p^0 = 2 p^z$.
The following on-shell relations greatly simplify the calculation:
\begin{align} \label{eq:projectors}
 \slashed{p} \, u_s(p) = \bar u_s(p) \, \slashed{p} = 0
 \,,\qquad
 P_n u_s(p) = u_s(p)
 \,,\qquad
 \bar u_s(p) P_\bn = \bar u_s(p)
\,,\end{align}
where $P_n = \slashed{n}\slashed{\bn}/4$ and $P_\bn = \slashed{\bn}\slashed{n}/4$ are projectors.
We employ Feynman gauge and regulate both infrared (IR) and UV divergences by working in $d=4-2\eps$ dimensions.
We note that in this setup, both the unpolarized TMDPDF and unpolarized quasi-TMDPDF already exist in the literature, which provides a useful reference for our analysis.

\paragraph{Notation.}
We work exclusively in Fourier space, with $b_\perp^\mu = (0,\bt,0)$ Fourier conjugate to $q_T^\mu = (0,\qt,0)$, and $b_\perp^2 = -\bt^2$.
We define a shorthand for the canonical logarithm that appears as
\begin{align}
 L_b = \ln\frac{\bt^2 \mu^2}{b_0^2} \,,\qquad b_0 = 2 e^{-\gamma_E}
\,,\end{align}
where $\gamma_E$ is the Euler-Mascheroni constant.
We express all our results in the $\MS$~scheme.
The associated renormalization scale $\mu$ is related to the MS scale $\mu_0$ by
$\mu^2 = 4\pi e^{-\gamma_E} \mu_0^2$.

\subsection{Tree-level results}
\label{sec:tree}

The tree-level results for the TMDPDF and quasi-TMDPDF are 
\begin{align} \label{eq:tree}
 f_{q/h_S}^{[\Gamma]\,(0)}(x,\bt,\mu,\zeta) &
 = \frac{1}{p^-} \Tr\biggl[u_s(p) \bar u_s(p) \frac{\Gamma}{2}\biggr] \delta(1-x)
\,,\nn\\
 \tilde f_{q/h_S}^{[\tilde\Gamma]\,(0)}(x,\bt,\mu,p^z) &
 = \frac{N_\Gamma}{p^z} \Tr\biggl[u_s(p) \bar u_s(p) \frac{\tilde\Gamma}{2}\biggr] \delta(1-x)
\,.\end{align}
For the case of a massless spinor $N_{\tilde\Gamma} = 1$ (i.e. independent of $\tilde\Gamma$),  the completeness relation reads:
\begin{align} \label{eq:completeness_massless}
 u_s(p) \bar u_s(p) = \frac12\,  \slashed{p}\, (1 - \Lambda \gamma_5 + \gamma_5 \slashed{s}_\perp)
\,,\end{align}
where $\Lambda$ is the helicity and $s_\perp$ is the transverse polarization vector.
(See \refcite{Barone:2001sp} for a review of polarization with spinors.)
Thus, the traces in \eq{tree} give
\begin{alignat}{3} \label{eq:traces_massless}
 &\Gamma = \gamma^\lambda: \qquad
 &&\Tr\Bigl[u_s(p) \bar u_s(p) \frac{\gamma^\lambda}{2}\Bigr]
 &&= p^\lambda
\,,\nn\\
 &\Gamma = \gamma^\lambda\gamma_5: \qquad
 &&\Tr\Bigl[u_s(p) \bar u_s(p) \frac{\gamma^\lambda \gamma_5}{2}\Bigr]
 &&= \Lambda p^\lambda
\,,\nn\\
 &\Gamma = \img \sigma^{\alpha \lambda} \gamma_5: \qquad
 &&\Tr\Bigl[u_s(p) \bar u_s(p) \frac{\img \sigma^{\alpha \lambda} \gamma_5}{2}\Bigr]
 &&= p^\lambda s_\perp^\alpha
\,.\end{alignat}
Comparing to the decompositions in \eqs{tmd_decomposition}{qtmd_decomposition}, we see that only the unpolarized ($f_1$ and $\tilde f_1$), helicity ($g_{1L}$ and $\tilde g_{1L}$) and transversity ($h_1$ and $\tilde h_1$) TMD and quasi-TMDs are nonzero at tree level. Moreover, by definition these functions are  normalized to $\delta(1-x)$ at tree level.
This implies that 
\begin{align}
C_{\rm ns}^{\tilde f_1} = C_{\rm ns}^{\tilde g_{1L}} 
  = C_{\rm ns}^{\tilde h_{1}}  =1 +{\cal O}(\alpha_s) \,.
\end{align}
It is instructive to carry out the same analysis using massive quarks.
In this case, the completeness relation is 
\begin{align} \label{eq:completeness_massive}
 u_s(p) \bar u_s(p) = (\slashed{p} + m) \frac12 (1 + \gamma_5 \slashed{s})
\,,\end{align}
and the Dirac traces in \eq{tree} evaluate to (in $d=4$ dimensions)
\begin{alignat}{3} \label{eq:traces_massive}
 &\Gamma = \gamma^\lambda: \qquad
 &&\Tr\Bigl[u_s(p) \bar u_s(p) \frac{\gamma^\lambda}{2}\Bigr]
 &&= p^\lambda
\,,\nn\\
 &\Gamma = \gamma^\lambda\gamma_5: \qquad
 &&\Tr\Bigl[u_s(p) \bar u_s(p) \frac{\gamma^\lambda \gamma_5}{2}\Bigr]
 &&= m s^\lambda
\,,\nn\\
 &\Gamma = \img \sigma^{\alpha \lambda} \gamma_5: \qquad
 &&\Tr\Bigl[u_s(p) \bar u_s(p) \frac{\img \sigma^{\alpha \lambda} \gamma_5}{2}\Bigr]
 &&= p^\lambda s^\alpha - p^\alpha s^\lambda  = p^\lambda s_\perp^\alpha
\,.\end{alignat}
In the last equation, we used that $p^\alpha = 0$ since $\alpha$ is a transverse index.
From \eq{spin_vector} $m s^0 = p^z$ and $m s^3 = p^0$, and using our
definition of $N_{\tilde \Gamma}$ in \eq{NGamma}, we find that both the TMD and the quasi-TMD are again normalized to $\delta(1-x)$ at tree level.
This result confirms the choices made in \eq{NGamma}.

\subsection{TMDPDFs}
\label{sec:tmd_nlo}
\FloatBarrier

To carry out the matching calculation to ${\cal O}(\alpha_s)$ requires comparing one-loop results for the spin-dependent TMDPDF and quasi-TMDPDF.
Here we present the one-loop results for the spin-dependent TMDPDF, which requires us to combine the spin-dependent beam function at one-loop with the standard TMD soft function at the same order.
Both the beam and the soft function are individually rapidity divergent, and we choose to employ the $\eta$ regulator of \refcites{Chiu:2011qc,Chiu:2012ir} to regulate these divergences.

\begin{figure}[t]
 \centering
 \begin{subfigure}{0.3\textwidth}
  \includegraphics[width=\textwidth]{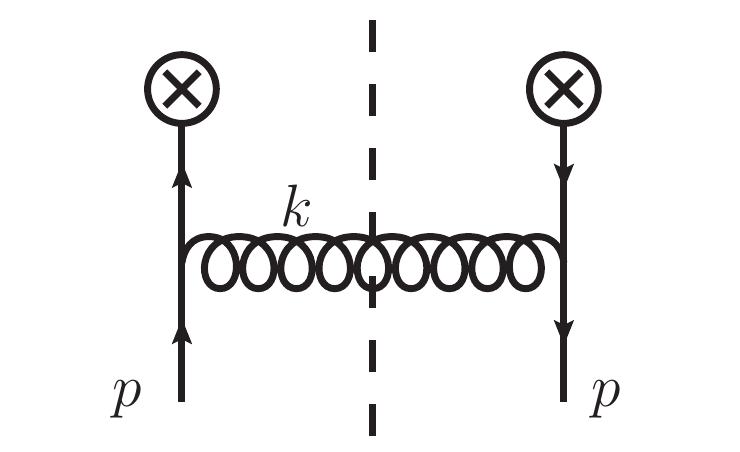}
  \caption{}
 \end{subfigure}
 \quad
 \begin{subfigure}{0.3\textwidth}
  \includegraphics[width=\textwidth]{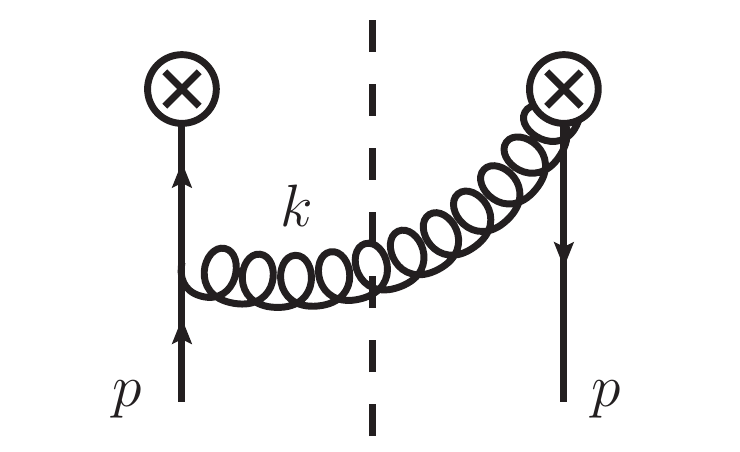}
  \caption{}
 \end{subfigure}
 \caption{One-loop diagrams contributing to the spin-dependent beam function  in \eq{beamfunc}.
   Mirror diagrams and the scaleless wave function and tadpole diagrams are not shown.
   The dashed line indicates the on-shell constraint on the emitted gluon.
   The $\otimes$ denotes the two Wilson line operators.
 }
 \label{fig:diagrams_tmdpdf}
\end{figure}

\paragraph{Spin-dependent beam function.}
There are only two diagrams that do not vanish in dimensional regularization, as shown in \fig{diagrams_tmdpdf}.
Note that in the lightlike case, the transverse Wilson line does not contribute in nonsingular gauges, such as the Feynman gauge we employ here.
Rapidity divergences for the phase space integral over $k$ are regulated by the factors of $|k^-/\nu|^{-\eta}$ appearing from the regulated Wilson lines, after which the calculation is straightforward.
We obtain
\begin{align} \label{eq:q_A}
 q_A^{\Gamma\,(1)}(x,\bt) &
 = \frac{1}{p^-} \bigl[\bar u_s^a(p) u_s^b(p)\bigr] \, \frac{\as C_F}{4\pi}
   \frac{\Gamma(-\eps)}{e^{\eps \gamma_E}} e^{\eps L_b}
   \Bigl(\gamma^\mu \gamma_\alpha \frac{\Gamma}{2} \gamma_\beta\gamma_\mu \Bigr)^{ab}
   \nn\\&~\quad\times
   \left[-\img \eps \frac{\bn^\alpha b_\perp^\beta + \bn^\beta b_\perp^\alpha}{ p^- b_\perp^2}
         + (1-x) \left(\frac{g_\perp^{\alpha\beta}}{2} + \eps \frac{b_\perp^\alpha b_\perp^\beta}{b_\perp^2}\right)
   \right]
\,,\\ \label{eq:q_B}
 q_B^{\Gamma\,(1)}(x,\bt) &
 = \frac{1}{p^-} \bigl[\bar u_s^a(p) u_s^b(p)\bigr]
   \frac{\as C_F}{4\pi} \frac{\Gamma(-\eps)}{e^{\eps \gamma_E}} e^{\eps L_b}
   \biggl\{
   - \bigl(P_\bn \Gamma + \Gamma P_n \bigr)^{ab}
   + \Bigl[\cL_0(1-x)
   \nn\\&~\quad
   + \delta(1-x) \Bigl( -\frac{1}{\eta} + \ln\frac{p^-}{\nu}\Bigr)\Bigr]
   \biggl[ \bigl( P_\bn \Gamma + \Gamma P_n \bigr)
   - \img \eps \frac{\slashed{\bn} \slashed{b}_\perp \Gamma}{p^- b_\perp^2}
   - \img \eps \frac{\Gamma \slashed{b}_\perp \slashed{\bn}}{p^- b_\perp^2}
   \biggr]^{ab} \biggr\}
\,,\end{align}
where $\cL_0(1-x) = [1/(1-x)]_+$ is the standard plus distribution.\footnote{Foreshadowing, note that we reproduce \eq{q_A} from the quasi beam function vertex diagram calculation in \sec{qtmd_nlo}.}
Since the graph leading to \eq{q_A} is rapidity finite it does not depend on the rapidity regulator.
In \eq{q_B} the rapidity divergence is explicit through the $1/\eta$ pole in the rapidity regulator $\eta$, with $\ln(p^-/\nu)$ the associated rapidity logarithm. Here, we have already expanded in $\eta\to0$, while keeping the exact $\eps$ dependence.

Both \eqs{q_A}{q_B} contain Dirac structures that scale as $1/(p^- b_\perp^2)$. We can neglect these terms in the limit of large $p^- \to \infty$, which is also why they do not appear in the decomposition in \eq{tmd_decomposition_2}.
Using \eq{projectors}, we obtain
\begin{align} \label{eq:B_nlo}
 B_{q/h_S}^{[\Gamma]}(x,\bt,\eps,p^-/\nu) &
 = \frac{1}{p^-}\bigl[\bar u_s^a(p) u_s^b(p)\bigr] \frac{\Gamma^{ab}}{2} \delta(1-x)
   +\frac{1}{p^-}\bigl[\bar u_s^a(p) u_s^b(p)\bigr] \, \frac{\as C_F}{4\pi}
   \frac{\Gamma(-\eps)}{e^{\eps \gamma_E}} e^{\eps L_b}
   \nn\\&\quad\times
   \biggl\{
   \frac{\Gamma^{ab}}{2} \Bigl[4 \cL_0(1-x) + 4 \delta(1-x) \Bigl( -\frac{1}{\eta} + \ln\frac{p^-}{\nu}\Bigr) - 4\Bigr]
   \nn\\&\qquad
   + \Bigl(\gamma^\mu \gamma_\alpha \frac{\Gamma}{2} \gamma_\beta\gamma_\mu \Bigr)^{ab}  (1-x)
   \biggl(\frac{g_\perp^{\alpha\beta}}{2} + \eps \frac{b_\perp^\alpha b_\perp^\beta}{b_\perp^2} \biggr)
   \biggr\} + \cO(\as^2)
\,.\end{align}

\paragraph{Soft function.}
From \refcites{Chiu:2012ir,Luebbert:2016itl} the bare soft function with the $\eta$ regulator is
\begin{align} \label{eq:S_nlo}
 S_q(b_T,\eps,\nu)
 = 1 + \frac{\as C_F}{4\pi} \frac{\Gamma(-\eps)}{e^{\eps \gamma_E}} e^{\eps L_b}
       \biggl[ \frac{8}{\eta} + 4 L_b + 8 \ln\frac{\nu}{\mu} - 4 \gamma_E - 4 \psi(-\eps) \biggr]
 + \cO(\as^2)
\,,\end{align}
where $\psi(x) = \Gamma'(x)/\Gamma(x)$ is the digamma function.

\paragraph{Spin-dependent TMDPDF.}
When using the $\eta$ regulator, the zero bin is scaleless and vanishes. Thus, the bare spin-dependent TMDPDF is
\begin{align} \label{eq:tmd_nlo}
 f_{q/h_S}^{[\Gamma]}(x,\bt,\eps,\zeta) &
 = B_{q/h_S}^{[\Gamma]}\bigl(x,\bt,\eps,\sqrt{\zeta}/\nu\bigr) \sqrt{S_q(b_T,\eps,\nu)}
\nn\\&
 = \frac{1}{p^-} \bigl[\bar u^a_s(p) u_s^b(p)\bigr] \frac{\Gamma^{ab}}{2} \delta(1-x)
   + \frac{1}{p^-} \bigl[\bar u_s^a(p) u_s^b(p)\bigr] \, \frac{\as C_F}{4\pi}
   \frac{\Gamma(-\eps)}{e^{\eps \gamma_E}} e^{\eps L_b}
   \nn\\&\quad\times
   \biggl\{
   \frac{\Gamma^{ab}}{2} \Bigl[ 4 \cL_0(1-x) + 2 \delta(1-x) \Bigl(L_b + \ln\frac{\zeta}{\mu^2} - \gamma_E - \psi_{-\eps} \Bigr) - 4\Bigr]
   \nn\\&\qquad
   + \Bigl(\gamma^\mu \gamma_\alpha \frac{\Gamma}{2} \gamma_\beta\gamma_\mu\Bigr)^{ab} (1-x)
      \biggl(\frac{g_\perp^{\alpha\beta}}{2} + \eps \frac{b_\perp^\alpha b_\perp^\beta}{b_\perp^2}\biggr)
   \biggr\}
 + \cO(\as^2)
\,.\end{align}
This result agrees with \refcite{Gutierrez-Reyes:2017glx} after accounting for different conventions.

It is convenient to note that \eq{tmd_nlo} has a universal structure because large parts of the NLO correction have the same Dirac structure as the tree level result.
To make this explicit, we write \eq{tmd_nlo} as
\begin{align} \label{eq:tmd_final}
 f_{q/h_S}^{[\Gamma]}(x, \bt, \eps, \zeta) &
 = \frac{1}{p^-} \Bigl[\bar u_s(p) \frac{\Gamma}{2} u_s(p)\Bigr]
   \Bigl[ \delta(1-x) + \frac{\as C_F}{4\pi} f_{b}^{(1)} \Bigr]
\nn\\&\quad
 + \frac{\as C_F}{4\pi} \frac{1}{p^-} \Bigl[\bar u_s(p) \gamma^\mu \gamma_\alpha \frac{\Gamma}{2} \gamma_\beta\gamma_\mu u_s(p)\Bigr]
   \biggl[ f^{(1)}_{a1} \frac{g_\perp^{\alpha\beta}}{2}
         + f^{(1)}_{a2} \Bigl( \frac{g_\perp^{\alpha\beta}}{2} - \frac{b_\perp^\alpha b_\perp^\beta}{b_\perp^2} \Bigr) \biggr]
\nn\\&\quad
 + \cO(\as^2)
\,,\end{align}
where the coefficient functions $f_i^{(1)}$, whose arguments we keep implicit, are given by
\begin{align} \label{eq:tmd_nlo_components}
 f^{(1)}_{b} &
 = \Bigl(\frac{1}{\eps_{\rm IR}} + L_b\Bigr) \bigl[ -2 P_{qq}(x) + 2 (1-x) \bigr]
 \nn\\&\quad
   + \delta(1-x) \biggl[ \frac{2}{\eps^2} + \Bigl(\frac{1}{\eps} + L_b\Bigr) \Bigl(-2 \ln\frac{\zeta}{\mu^2} + 3\Bigr) - L_b^2 - \frac{\pi^2}{6} \biggr] + \cO(\eps)
\,,\nn\\
 f^{(1)}_{a1} &= -(1-x) \Bigl[\frac{1}{\eps_{\rm IR}} + L_b + 1 + \cO(\eps) \Bigr]
\,,\nn\\
 f^{(1)}_{a2} &= (1-x)  + \cO(\eps)
\,.\end{align}
Here, $\eps_{\rm IR}$ denotes poles of IR origin, while $\eps$ denotes poles of UV origin.
The unpolarized splitting function is given by
\begin{align}
 P_{qq}(x) = \biggl(\frac{1+x^2}{1-x}\biggr)_+
 = 2 \cL_0(1-x) + \frac32 \delta(1-x) - (1+x)
\,.\end{align}
In \eq{tmd_nlo_components}, 
$f_b^{(1)}$ arises entirely from diagram (b) and the soft subtraction, whereas the two distinct Lorentz structures of $f_{a1}^{(1)}$ and $f_{a2}^{(1)}$ result from diagram (a).
\Eqs{tmd_final}{tmd_nlo_components} reveal that at NLO, the full distributional structure in $x$ of the spin-dependent TMDPDF is proportional to the tree-level normalization and is thus universal.  The true dependence on $\Gamma$ is always proportional to $(1-x)$.
This also explains the similar kernels that appear when perturbatively matching the spin-dependent TMD onto spin-dependent PDFs, see \refcites{Bacchetta:2013pqa,Echevarria:2015uaa,Gutierrez-Reyes:2017glx}.

For any desired Dirac structure $\Gamma$, it is straightforward to insert this $\Gamma$ into \eq{tmd_nlo_components} and expand in $\eps$ to obtain the  various desired spin-dependent TMDPDFs.
For the three Dirac structures we consider, we obtain the following renormalized results:
\begin{align} \label{eq:tmd_renorm}
 f_{q/h_S}^{[\gamma^\lambda]}(x, \bt, \mu, \zeta) &
 = \delta(1-x) \biggl\{ 1 + \frac{\as C_F}{4\pi} \biggl[-L_b^2 + L_b \Bigl(2 \ln\frac{\mu^2}{\zeta} + 3\Bigr)  - \frac{\pi^2}{6} \biggr] \biggr\}
 \nn\\&\quad
 + \frac{\as C_F}{4\pi} \biggl[ -2 \Bigl(\frac{1}{\eps_{\rm IR}} + L_b\Bigr) P_{qq}(x) + 2 (1-x) \biggr]
 + \cO(\as^2)
\,,\nn\\
 f_{q/h_S}^{[\gamma^\lambda\gamma_5]}(x, \bt, \mu, \zeta) &
 = \Lambda  f_{q/h_S}^{[\gamma^\lambda]}(x, \bt, \mu, \zeta)
\,,\nn\\
 f_{q/h_S}^{[\img \sigma^{\alpha-}\gamma_5]}(x, \bt, \mu, \zeta) &
 = s_\perp^\alpha \delta(1-x) \biggl\{ 1 + \frac{\as C_F}{4\pi} \biggl[-L_b^2 + L_b \Bigl(2 \ln\frac{\mu^2}{\zeta} + 3\Bigr)  - \frac{\pi^2}{6} \biggr] \biggr\}
 \nn\\&\quad
 + s_\perp^\alpha\frac{\as C_F}{4\pi} \biggl[ -2 \Bigl(\frac{1}{\eps_{\rm IR}} + L_b\Bigr) P^T_{qq}(x) \biggr]
 + \cO(\as^2)
\,,\end{align}
where in the second equation we use that at one loop the unpolarized and helicity splitting functions agree, whereas the transverse splitting function is given by~\cite{Artru:1989zv}
\begin{align}
 P^T_{qq}(x) &
 = P_{qq}(x) - (1-x)
 = 2 \cL_0(1-x) - 2 + \frac32 \delta(1-x)
\,.\end{align}
In \eq{tmd_renorm}, the $\delta(1-x)$ terms agree between all three results because they arise solely from the sail diagram, which is proportional to the tree-level Dirac structure and has the same value for different choices of $\Gamma$.
The nontrivial $x$ dependence differs, as it obtains contributions from the vertex diagram, which agrees between the unpolarized and helicity structures, but vanishes for the transverse structure.

Finally, we briefly comment in more detail on the calculation of the leading spin-dependent TMDs in \refcite{Gutierrez-Reyes:2017glx}, where a similar result of the combination of \eqs{q_A}{q_B} was presented, but with bare results employing the $\delta$-rapidity regulator. See Eq.~(14) of \refcite{Gutierrez-Reyes:2017glx} for the unsubtracted bare beam function.
(We remark that their result contains an additional imaginary term proportional to $\img\pi/2$, which does not contribute to their final result.) 
We also note that a different argument to eliminate the $1/(p^- b_T^2)$ suppressed terms in \eq{q_B} is given in \refcite{Gutierrez-Reyes:2017glx} (the corresponding term in \eq{q_A} is not present in their result):
by choosing a scheme that imposes the condition $\slashed{\bn} \Gamma = \Gamma \slashed{\bn}$, these terms immediately vanish, as required to cancel the rapidity divergences.
It is then observed that the Dirac structures $\Gamma$ allowed by this criterion are precisely those of leading power. This is consistent with our observation that due to the $1/p^-$, these terms only contribute at higher orders in the power expansion.%
\footnote{There appears to be a typo in \refcite{Gutierrez-Reyes:2017glx} where the $1/p^-$ factor is missing in their result, such that the designation of this term as power suppressed is not obvious. It is easy to see that this factor is required to obtain the correction mass dimension of their $\slashed{b}_\perp / b_\perp^2$ terms.}
A key advantage of our observation compared to \refcite{Gutierrez-Reyes:2017glx} is that by immediately discarding power suppressed terms, we do not need to restrict the allowed schemes for treating $\gamma^5$ in $d$ dimensions (as was done in~\cite{Gutierrez-Reyes:2017glx} with the Larin$^+$ scheme).
This provides evidence for a more intricate structure of rapidity divergences at higher order in the power expansion that are not canceled by the leading power soft function alone.

\subsection{Quasi-TMDPDFs}
\label{sec:qtmd_nlo}

In this section, we calculate the quasi-TMDPDFs for various choices of spin polarizations.  The calculation for the unpolarized quasi beam function can be found in \refcite{Ebert:2019okf}, which provides a baseline for our analysis.   \refcite{Ebert:2019okf} keeps intermediate results exact as long as possible, with terms suppressed by $1/L$ or $1/(b_T P^z)$ only dropped at the end to extract the leading-power matching. 
Here, we instead carry out these expansions earlier on. As we see below, this allows us to relate the spin-dependent beam function to the unpolarized case, thus rendering many explicit calculations unnecessary.

\begin{figure*}
 \centering
 \begin{subfigure}{0.3\textwidth}
  \includegraphics[width=\textwidth]{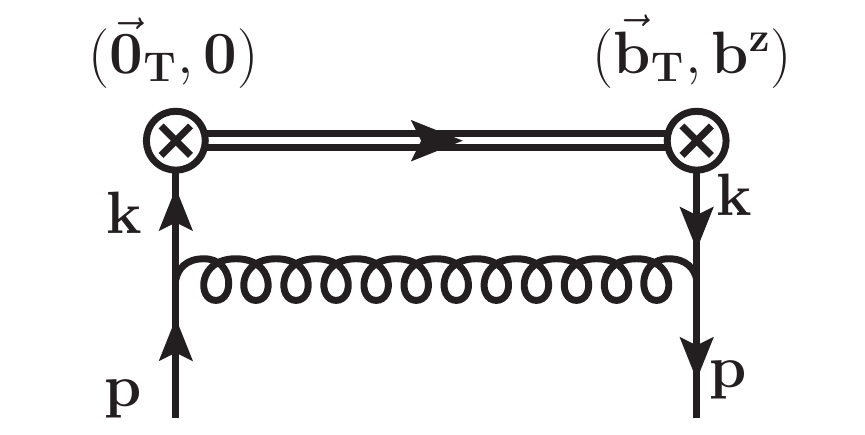}
  \caption{Vertex diagram}
  \label{fig:qTMD_a}
 \end{subfigure}
 \quad
 \begin{subfigure}{0.3\textwidth}
  \includegraphics[width=\textwidth]{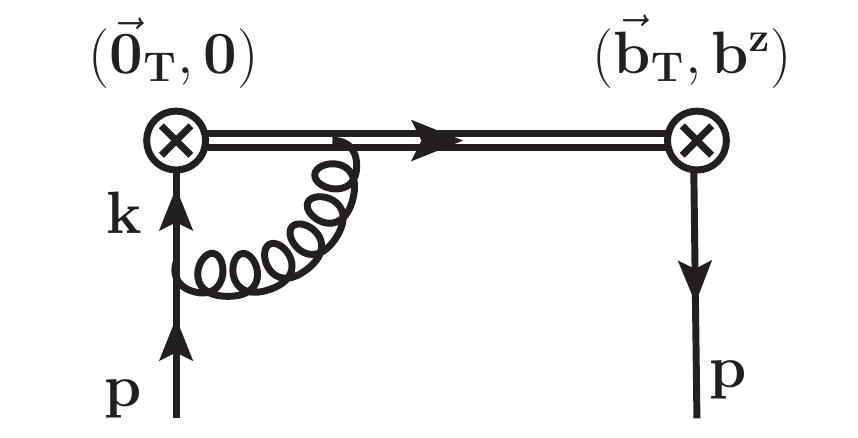}
  \caption{Sail topology}
  \label{fig:qTMD_b}
 \end{subfigure}
 \quad
 \begin{subfigure}{0.3\textwidth}
  \includegraphics[width=\textwidth]{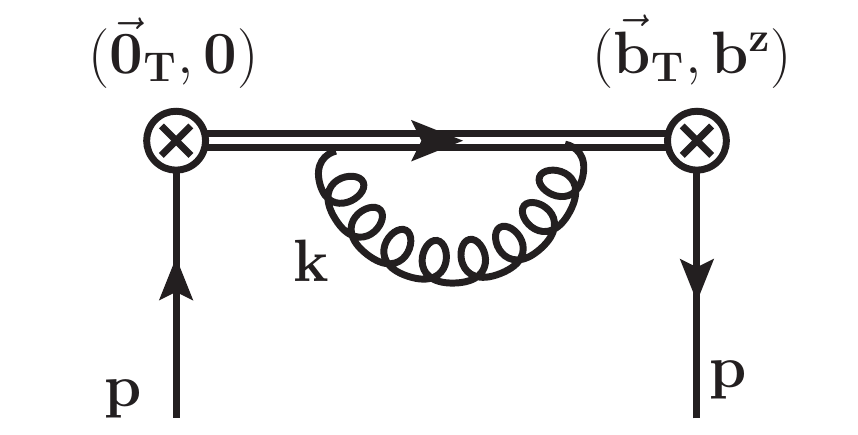}
  \caption{Tadpole}
  \label{fig:qTMD_c}
 \end{subfigure}
 \caption{One-loop diagrams contributing to the quasi-TMD beam function of \eq{qbeam_a}
          in Feynman gauge. Mirror diagrams are not shown.
          The double line represents the Wilson line $W_{\sqsubset}$. 
 }
 \label{fig:qbeamfunc_nlo}
\end{figure*}

Three types of diagrams contribute in Feynman gauge: the vertex, tadpole, and sail topologies shown in \fig{qbeamfunc_nlo}. We start our analysis of each diagram in coordinate space.
We discuss each of these cases separately.

\subsubsection*{Vertex diagram}

The vertex diagram in \fig{qTMD_a} gives \cite{Ebert:2019okf}
\begin{align} \label{eq:qtmd_a_1}
 \tilde q_A^{\tilde\Gamma\,(1)}(b, p) &
 = \bigl[ \bar u^a_s(p) u_s^b(p)\bigr] \frac{\as C_F}{4\pi}
   \Bigl(\gamma^\mu \gamma_\alpha \frac{\tilde\Gamma}{2} \gamma_\beta \gamma_\mu \Bigr)^{ab}
   \times \cI_v^{\alpha\beta}(b,p)
\,,
\end{align}
where the integral $\cI_v^{\alpha\beta}$ is given by
\begin{align}
  \cI_v^{\alpha\beta}(b,p) &= -4 \img \mu_0^{2\eps} \int\frac{\df^dk}{(2\pi)^{d-2}} \frac{k^\alpha k^\beta}{k^4 (p-k)^2} 
  \, e^{\img k \cdot b}
\,.\end{align}
To evaluate \eq{qtmd_a_1}, we first evaluate
\begin{align} \label{eq:Iv_1}
 \cI_v(b^2, p \cdot b) &
 = -4 \img \mu_0^{2\eps} \int\frac{\df^dk}{(2\pi)^{d-2}} \frac{e^{\img k \cdot b}}{k^4 (p-k)^2}
 \nn\\&
 = -\frac14 (\pi \mu_0^2)^\eps \Gamma(-1-\eps) \frac{1 + \img \pb - e^{\img \pb}}{(\pb)^2} (-b^2)^{1+\eps}
\,,\end{align}
which can be derived using standard techniques.
Note that here we have made explicit that the integral can only depend on the two Lorentz invariants $b^2$ and $\pb$, as $p^2 = 0$.
From here, we note that
\begin{align} \label{eq:Iv_2}
 \cI_v^{\alpha\beta}(b, p) &
 = \left(\img\frac{\partial}{\partial b_\alpha}\right) \left(\img\frac{\partial}{\partial b_\beta}\right)  \cI_v(b^2,\pb)
 \nn\\ &
 = -2 g^{\alpha\beta} \frac{\partial \cI_v}{\partial (b^2)}
 - 2 (p^\alpha b^\beta + b^\alpha p^\beta) \frac{\partial^2 \cI_v}{\partial (b^2) \partial (\pb)}
 - 4 b^\alpha b^\beta \frac{\partial^2 \cI_v}{\partial^2 (b^2)}
 - p^\alpha p^\beta \frac{\partial^2 \cI_v}{\partial^2(\pb)}
\nn\\&
 = -2 g_\perp^{\alpha\beta} \frac{\partial \cI_v}{\partial (b^2)}
 - 4 b^\alpha b^\beta \frac{\partial^2 \cI_v}{\partial^2 (b^2)}
 + \cO(n^\alpha, n^\beta)
\,.\end{align}
When plugging \eq{Iv_2} into \eq{qtmd_a_1}, all terms with $n^\alpha$ or $n^\beta$ (or equivalently $p^\alpha$ or $p^\beta$) vanish due to $\slashed{n} u_s(p) = 0$.
This allows us to replace the full metric $g$ with the transverse metric $g_\perp$.
It is now straightforward to evaluate \eq{qtmd_a_1},
\begin{align} \label{eq:qtmd_a_2}
 \tilde q_A^{\tilde\Gamma\,(1)}(b, p) &
 = \bigl[ \bar u^a_s(p) u_s^b(p)\bigr] \frac{\as C_F}{4\pi}
   \frac{\Gamma(-\eps)}{e^{\eps\gamma_E}} e^{\eps L_b}
   \Bigl(\gamma^\mu \gamma_\alpha \frac{\tilde\Gamma}{2} \gamma_\beta \gamma_\mu \Bigr)^{ab}
   \nn\\&\quad\times
   \biggl(\frac{g_\perp^{\alpha\beta}}{2} + \eps \frac{b^\alpha b^\beta}{b^2} \biggr)
   \biggl(\frac{-b^2}{b_T^2}\biggr)^{\eps} \frac{1 + \img \pb - e^{\img \pb}}{(\pb)^2}
\,.\end{align}
We write \Eq{qtmd_a_2} in terms of the Lorentz invariants $b^2$ and $p\cdot b$ so that it can be used to obtain both the TMDPDF and quasi-TMDPDF vertex diagrams. 
For the TMD, $\pb = \frac12 p^- b^+$ and $b^2 = b_\perp^2 = -b_T^2$,
whereas for the quasi-TMD we have $\pb = -p^z b^z$ and $b^2 = -b_T^2 - (b^z)^2$.
In both cases, the required Fourier transform can be obtained from
\begin{align} \label{eq:int_vertex}
 \int\frac{\df (\pb)}{2\pi} e^{-\img x \pb} \, (\pb)^n \, \frac{1 + \img \pb - e^{\img \pb}}{(\pb)^2}
 = \img^n \frac{\df^n}{\df^n x} (1-x) \theta(x) \theta(1-x)
\,.\end{align}
Note that the physical support $0 \le x \le 1$ naturally arises in the integral in \eq{int_vertex}.  In the following we explicitly suppress the $\theta$ functions.
When expanding \eq{qtmd_a_2} in small $(b^z/b_T)^2$, only the $n=0$ term contributes, yielding
\begin{align} \label{eq:qtmd_a_final}
 \tilde q_A^{\tilde\Gamma\,(1)}(b, p) &
 = \frac{1}{p^z}\bigl[ \bar u^a_s(p) u_s^b(p)\bigr] \frac{\as C_F}{4\pi}
   \frac{\Gamma(-\eps)}{e^{\eps\gamma_E}} e^{\eps L_b}
   \Bigl(\gamma^\mu \gamma_\alpha \frac{\tilde\Gamma}{2} \gamma_\beta \gamma_\mu \Bigr)^{ab}
   (1-x) \left(\frac{g_\perp^{\alpha\beta}}{2} + \eps \frac{b_\perp^\alpha b_\perp^\beta}{b_\perp^2} \right)
\nn\\&
 = \frac{1}{p^z}\bigl[ \bar u^a_s(p) u_s^b(p)\bigr] \frac{\as C_F}{4\pi}
   \Bigl(\gamma^\mu \gamma_\alpha \frac{\tilde\Gamma}{2} \gamma_\beta \gamma_\mu \Bigr)^{ab}
   \nn\\&\quad\times
   (x-1) \left[ \frac{g_\perp^{\alpha\beta}}{2} \left(\frac{1}{\eps} + L_b + 1\right)
   - \left( \frac{g_\perp^{\alpha\beta}}{2} - \frac{b_\perp^\alpha b_\perp^\beta}{b_\perp^2}\right)
   \right]
\,.\end{align}
In the same fashion, we can obtain the vertex diagram of the TMDPDF, given above in \eq{q_A}, where the $b_T/p^-$-suppressed term has $n=1$ and can be neglected at leading power.
The close relationship between the two results is not surprising, as the two calculations only differ by the definition of $\tilde\Gamma$ and the choice of $b^\mu$.

\subsubsection*{Sail diagram}

Next we consider the sail diagram in \fig{qTMD_b} and its mirror diagram.
Extracting the overall Dirac structure, we write the sail diagram as
\begin{align} \label{eq:qtmd_b_1}
 \tilde q_B^{\tilde\Gamma\,(1)}(b) &
 =  \frac{\as C_F}{2 \pi} \Bigl\{
     \bigl[ \bar u_s(p) \tilde\Gamma \gamma^\rho \gamma^\mu u_s(p) \bigr] \tilde q^{\tilde\Gamma\,(B1)}_{\mu\rho}(b^\mu)
   + \bigl[ \bar u_s(p) \gamma^\mu \gamma^\rho \tilde\Gamma u_s(p) \bigr] \tilde q^{\tilde\Gamma\,(B2)}_{\mu\rho}(b^\mu) \Bigr\}
\,,\nn\\
 \tilde q^{\tilde\Gamma\,(B1)}_{\mu\rho}(b) &
 = - \frac{\mu_0^{2\eps}}{(2\pi)^{d-2}} \int\!\df^d k \int_0^1 \df s \,
  \frac{\gamma'(s)_\mu k_\rho}{k^2 (p-k)^2} \, e^{\img p \cdot b -\img (p-k) \cdot \gamma(s)}
\,,\nn\\
 \tilde q^{\tilde\Gamma\,(B2)}_{\mu\rho}(b) &
 = - \frac{\mu_0^{2\eps}}{(2\pi)^{d-2}} \int\!\df^d k \int_0^1 \df s \,
  \frac{\gamma'(s)_\mu k_\rho}{k^2 (p-k)^2} \, e^{\img k \cdot b + \img (p-k)\cdot\gamma(s)}
\,,\end{align}
where $\gamma(s)$ is the path of the Wilson lines.
To make a connection to the calculation presented in \refcite{Ebert:2019okf}, we work with the path given in \eq{qbeam_a} and illustrated in the left panel of \fig{qwilsonlines}, but we stress that in the limit $L \gg b_T$ this path is equivalent to the one in \eq{qbeam_b}.

We insert $\tilde q^{\tilde\Gamma\,(B1)}_{\mu\rho}$ into $\tilde q_B^{\tilde\Gamma\,(1)}$ and simplify the result using the on-shell condition $\slashed{n} u_s(p) = 0$, giving
\begin{align} \label{eq:sail_dirac_1}
 \slashed{k} \slashed{\gamma}' u_s(p) &
 = \Bigl[ k^- \frac{\slashed{n}}{2} + k^+ \frac{\slashed{\bn}}{2} + \slashed{k}_\perp \Bigr]
   \Bigl[ \gamma^{\prime-} \frac{\slashed{n}}{2} + \gamma^{\prime+} \frac{\slashed{\bn}}{2} + \slashed{\gamma}'_\perp \Bigr]  u_s(p)
\nn\\&
 = \Bigl[ k^- \gamma^{\prime+} + \frac{\slashed{\bn}}{2} \bigl(k^+  \slashed{\gamma}'_\perp - \slashed{k}_\perp \gamma^{\prime+} \bigr)  + \slashed{k}_\perp \slashed{\gamma}'_\perp \Bigr] u_s(p)
\,.\end{align}
Here, we suppress the argument of $\gamma^\prime \equiv \gamma^\prime(s)$ and employ lightcone coordinates to make the on-shell condition manifest, noting that in practice $\gamma^{\prime+} = -\gamma^{\prime z}$.
The calculation for $\tilde q^{\tilde\Gamma\,(B2)}_{\mu\rho}(b^\mu)$ works similarly.

In \eq{sail_dirac_1}, the contributions with $\slashed{\gamma}_\perp'$ arise from the transverse gauge links.
In particular, evaluation of the line integrals in \eq{qtmd_b_1} gives
\begin{align}
 \int_0^1\!\! \df s \,  \gamma'(s)_\mu e^{\img q \cdot \gamma(s)} &
 = \img \hat z_\mu \frac{e^{-\img q^z L} - 1}{q^z}
 + \img b_{\perp\mu} e^{-\img q^z L} \frac{e^{-\img \bt \cdot \qt} - 1}{\bt \cdot \qt}
 + \img \hat z_\mu \frac{e^{- \img b^z q^z} - e^{- \img q^z L}}{q^z} e^{-\img \bt \cdot \qt}
\,.\end{align}
The perpendicular contribution contains a pure phase $e^{-\img q^z L}$, which oscillates quickly in the limit $L,p^z \to \infty$ and makes this term neglectable.
This is analogous to how transverse gauge links do not contribute in nonsingular gauges.
Thus, we can approximate \eq{sail_dirac_1}:
\begin{align} \label{eq:sail_dirac_2}
 \slashed{k} \slashed{\gamma}' u_s(p) &
 = \bigl( k^-  - \slashed{k}_\perp \bigr) \gamma^{\prime+} u_s(p)
 + \cO\Bigl(\frac{1}{p^z L}\Bigr)
\,.\end{align}
After Fourier-transforming \eq{qtmd_b_1} with respect to $b^z$, $\tilde q_B^{\tilde\Gamma\,(1)}$ only depends on $p^-$ and $b_\perp$.
Thus, the contributions from the Dirac structures $k^-$ and $\slashed{k}_\perp$ in \eq{sail_dirac_2} parametrically behave as $p^-$ and $\slashed{b}_\perp / b_T^2$, respectively.
The latter is suppressed in the limit $b_T p^z \gg 1$, similar to the subleading-twist terms in the physical TMDPDF, cf.~\eq{q_B}. We can thus further expand
\begin{align} \label{eq:sail_dirac_3}
 \slashed{k} \slashed{\gamma}' u_s(p) &
 = k^- \gamma^{\prime+} u_s(p) + \cO\Bigl(\frac{1}{p^z L}, \frac{1}{b_T p^z} \Bigr)
\,.\end{align}
Overall, in the physical limit \eq{qtmd_b_1} reduces to the tree-level Dirac structure, leaving us free to write
\begin{align} \label{eq:qtmd_b_final}
 \tilde q_B^{\tilde\Gamma\,(1)}(x,\bt) &
 = \frac{1}{p^z} \Bigl[\bar u_s(p) \frac{\tilde\Gamma}{2} u_s(p) \Bigr]
   \frac{\as C_F}{4\pi} \, \tilde q_B(x,\bt)
\,,\end{align}
where $\tilde q_B(x,\bt)$ is the one-loop coefficient of the unpolarized quasi beam function, i.e.,
\begin{align}
 \frac{1}{2} \sum_s \tilde q_B^{\gamma^z\,(1)}(x,\bt) = \frac{\as C_F}{4\pi} \, \tilde q_B(x,\bt)
\,.\end{align}
We can also read off from the result in \refcite{Ebert:2019okf} that
\begin{align} \label{eq:qtmd_b_unpol}
 \tilde q_B(x,\bt) &
 = \Bigl(\frac{1}{\eps} + L_b\Bigr) \Bigl[ -4 \cL_0(1-x) -2 \delta(1-x) L_{p^z}+ 4 \Bigr]
\nn\\&\quad
 + \delta(1-x) \Bigl(-\frac{2}{\eps} + \frac{2}{\eps}L_{p^z} - L_b^2 - L_{p^z}^2 + 2 L_{p^z} - 4 \Bigr)
\,,\end{align}
where $L_{p^z} = \ln[(2 p^z/\mu)^2]$. The plus distribution ${\cal L}_0(1-x)$ implicitly only has support for $x \in [0,1]$.
Note that the pole in the first line in \eq{qtmd_b_unpol} has an IR origin, and thus must agree with the corresponding IR pole in the TMDPDF so that the IR divergences cancel each other out in the matching.
By comparing to \eq{tmd_final}, it is easy to see that this holds upon identifying $\zeta = (2 p^z)^2$.

\subsubsection*{Tadpole diagram}

The Wilson line self-energy in \fig{qTMD_c} is given by
\begin{align} \label{eq:qtmd_c_final}
 \tilde q_C^{\tilde\Gamma\,(1)}(x,\bt) &
 = \frac{\img}{p^z} \Bigl[ \bar u_s(p) \frac{\tilde\Gamma}{2} u_s(p) \Bigr]
   \frac{\as C_F}{2\pi}  \frac{\mu_0^{2\eps}}{(2\pi)^{d-2}}
   \int_0^1 \df s \, \df t \, [\gamma'(s) \cdot \gamma'(t)]
   \int\!\df^d k \frac{e^{\img k \cdot [\gamma(s) - \gamma(t)]}}{k^2 + \img 0}
\nn\\&
 \equiv \frac{1}{p^z} \Bigl[\bar u_s(p) \frac{\tilde\Gamma}{2} u_s(p) \Bigr]
   \frac{\as C_F}{4\pi} \, \tilde q_C(x,\bt)
\,.\end{align}
Since the Wilson line vertices do not have a Dirac structure, the tadpole contribution is trivially proportional to the tree level Dirac structure.
In \eq{qtmd_c_final}, $\tilde q_C(x,\bt)$ is the coefficient of the tadpole diagram of the unpolarized quasi beam function because
\begin{align}
 \frac{1}{2} \sum_s \tilde q_C^{\gamma^z\,(1)}(x,\bt) = \frac{\as C_F}{4\pi} \, \tilde q_C(x,\bt)
\,.\end{align}
From \refcite{Ebert:2019okf} we can read off
\begin{align} \label{eq:qtmd_c_unpol}
 \tilde q_C(x,\bt) &= \delta(1-x) \biggl[ \frac{6}{\eps} + 6 L_b  + 4 + \frac{4\pi L}{b_T} + \cO(\eps) \biggr]
\,,\end{align}
up to corrections suppressed in the $P^z b_T\gg 1$ and $L/ b_T\gg 1$ limits.
We remark upon the explicit $L/b_T$ divergence in \eq{qtmd_c_unpol}, which must be canceled by the soft factor in a manner analogous to the cancellation of rapidity divergences that appears in the TMD case (see \refcite{Ebert:2019okf}).

\subsubsection*{Combined result}

The sum of the results for the vertex, sail and tadpole diagrams gives the bare quasi beam function at one loop,
\begin{align} \label{eq:qbeam_final}
 \tilde B_{q/h_S}^{[\tilde\Gamma]}(x, \bt, \eps, p^z) &
 = \frac{1}{p^z} \biggl[\bar u_s(p) \frac{\tilde\Gamma}{2} u_s(p) \biggr]
   \Bigl\{ \delta(1-x) + \frac{\as C_F}{4\pi} \bigl[\tilde B_b^{(1)} + \tilde R_B^{(1)} \delta(1-x) \bigr] \Bigr\}
\nn\\&\quad
 + \frac{\as C_F}{4\pi} \frac{1}{p^z} \Bigl[\bar u_s(p) \gamma^\mu \gamma_\alpha \frac{\tilde\Gamma}{2} \gamma_\beta \gamma_\mu u_s(p) \Bigr]
   \biggl[ \tilde B_{a1}^{(1)} \frac{g_\perp^{\alpha\beta}}{2}
         + \tilde B_{a2}^{(1)} \Bigl( \frac{g_\perp^{\alpha\beta}}{2} - \frac{b_\perp^\alpha b_\perp^\beta}{b_\perp^2} \Bigr) \biggr]
\nn\\&\quad
 + \cO(\as^2)
\,.\end{align}
The coefficient functions, whose arguments we keep implicit, are given by
\begin{align} \label{eq:qbeam_nlo_components}
 \tilde B_b^{(1)} &
 = \Bigl(\frac{1}{\eps_{\rm IR}} + L_b\Bigr) [-2 P_{qq}(x) + 2(1-x)]
\nn\\&\quad
 + \delta(1-x) \Bigl[ \frac{7}{\eps} - (L_b +  L_{P^z})^2  + 2 L_{P^z} + 9 L_b \Bigr] + \cO(\eps)
\,,\nn\\
 \tilde B_{a1}^{(1)} &= - (1-x)\Bigl(\frac{1}{\eps_{\rm IR}} + L_b + 1\Bigr) + \cO(\eps)
\,,\nn\\
 \tilde B_{a2}^{(1)} &= (1-x) + \cO(\eps)
\,.\end{align}
Note that we separated out the divergent term into $ \tilde R_B^{(1)} = 4\pi L / b_T$ and made the origin of all poles as either IR or UV explicit. 
Evaluating \eq{qbeam_nlo_components} for the three Dirac structures in \eq{qGamma}, we obtain the UV-renormalized quasi beam functions:
\begin{align} \label{eq:qbeam_renorm}
 \tilde B_{q/h_S}^{[\gamma^\lambda]}(x, \bt, \mu, p^z) &
 = \delta(1-x) \Bigl\{1 +  \frac{\as C_F}{4\pi} \Bigl[- (L_b +  L_{P^z})^2  + 2 L_{P^z} + 9 L_b + \tilde R_B^{(1)} \Bigr] \Bigr\}
 \nn\\&\quad
 + \frac{\as C_F}{4\pi} \Bigl[ - 2 \Bigl(\frac{1}{\eps_{\rm IR}} + L_b\Bigr) P_{qq}(x) + 2 (1-x) \Bigr]
 + \cO(\as^2)
\,,\nn\\
 \tilde B_{q/h_S}^{[\gamma^\lambda\gamma]}(x, \bt, \mu, p^z) &
 = \Lambda \tilde B_{q/h_S}^{[\gamma^\lambda]}(x, \bt, \mu, p^z) + \cO(\as^2)
\,,\nn\\
 \tilde B_{q/h_S}^{[\img \sigma^{\alpha \lambda} \gamma_5]}(x, \bt, \mu, p^z) &
 = s_\perp^\alpha \delta(1-x) \Bigl\{1 +  \frac{\as C_F}{4\pi} \Bigl[- (L_b +  L_{P^z})^2  + 2 L_{P^z} + 9 L_b + \tilde R_B^{(1)} \Bigr] \Bigr\}
 \nn\\&\quad
 + s_\perp^\alpha \frac{\as C_F}{4\pi} \Bigl[ - 2 \Bigl(\frac{1}{\eps_{\rm IR}} + L_b\Bigr) P^T_{qq}(x) \Bigr]
 + \cO(\as^2)
\,.\end{align}
Similar to \eq{tmd_renorm}, the $\delta(1-x)$ terms in \eq{qbeam_renorm} agree for all choices of $\tilde\Gamma$ as they arise entirely from the sail diagrams, which are proportional to the tree-level Dirac structure. We have explicitly validated this in \app{sail}.
The differences result entirely from the vertex diagram, which is absent for the transverse structure, see also \app{vertex}.
We also note that all results are independent of the choice $\lambda=t$ or $\lambda=z$.

\subsection{One-loop matching}

By comparing the results in \eqs{tmd_renorm}{qbeam_renorm} and noting that the unspecified quasi soft factor is spin independent, we see that the ratios of quasi-TMDPDFs and TMDPDFs agree for the unpolarized, helicity and transversity structures,
\begin{align} \label{eq:ratios_B_f}
   \frac{\tilde f_1(x, b_T, \mu, P^z)}{f_1(x, b_T, \mu, \zeta)} &
 = \frac{\tilde g_{1L}(x, b_T, \mu, P^z)}{g_{1L}(x, b_T, \mu, \zeta)}
 = \frac{\tilde h_1(x, b_T, \mu, P^z)}{h_1(x, b_T, \mu, \zeta)}
\,.\end{align}
From \eq{relation_ratio}, it follows that the short-distance matching kernel is identical for these functions. Using the result from \refcite{Ebert:2019okf}, we thus obtain
\begin{align}  \label{eq:Cnsall}
 C_\ns^{f_1}(\mu, x P^z) &
 = C_\ns^{g_{1L}}(\mu, x P^z)
 = C_\ns^{h_1}(\mu, x P^z)
 \nn\\&
 = 1 + \frac{\as C_F}{4\pi} \biggl[ - \ln^2\frac{(2 x P^z)^2}{\mu^2} + 2 \ln\frac{(2 x P^z)^2}{\mu^2}  - 4 + \frac{\pi^2}{6} \biggr] + \cO(\as^2)
\,.\end{align}
This is the main result of our analysis.  We discuss a reason for this observed universality in \sec{allordermatch}, where we also discuss the extension of this observation to higher orders.

It is interesting to note that the one-loop results of the both the TMDPDF and the quasi beam functions, see \eqs{tmd_final}{qbeam_final}, consist of universal kinematic structures, whereas the dependence on $\Gamma$ and $\tilde\Gamma$ only arises through the Dirac traces.
The Dirac traces are identical for each choice for $\Gamma$/$\tilde\Gamma$, and thus the spin independence of the ratios of quasi-TMDs to TMDs can be traced back to the universal kinematic structure of the one-loop diagrams.
Based on this observation, we present more general arguments for the spin independence of the matching kernel in \sec{allordermatch}.
In addition, since the tree-level Dirac structures are all normalized, only the Dirac trace arising from the vertex diagram can lead to differences between different spin-dependent quasi-TMDs, or between different spin-dependent TMDs, respectively.
Since the vertex diagram only encodes collinear interactions between the external quarks, which are not affected by the different Wilson line geometries, it obeys the simple boost picture underlying LaMET and does not affect the matching.
In contrast, the sail and tadpole diagrams, as well as the soft subtraction, resolve the Wilson line structure; thus, they induce a nontrivial matching.

The $f_{1T}^\perp$, $g_{1T}$, $h_{1L}^\perp$, and $h_{1}^\perp$ distributions, which are proportional to $b_\perp^\alpha / b_T$, cannot be constrained by the setup used for our calculation. The chosen on-shell quark state for the one-loop calculation has no overlap with the corresponding (quasi-)TMDPDF operators. The analysis for these functions is beyond the scope of the present work.

A special case is the pretzelosity distribution $h_{1T}^\perp$, which is proportional to $g_\perp^{\alpha\beta}/2 - b_\perp^\alpha b_\perp^\beta / b_\perp^2$, cf. \eq{tmd_decomposition_2}.
From \eqs{tmd_final}{qbeam_final}, it is clear that this contribution arises first at one loop from the vertex diagram for both the quasi and non-quasi TMDs.  It does not contain any divergences or logarithms and in both cases is simply proportional to
\begin{align} \label{eq:pretzelosity_trace}
 \Bigl[\bar u_s(p) \gamma^\mu \gamma_\alpha \frac{\img \sigma^{\rho\lambda} \gamma_5}{2} \gamma_\beta \gamma_\mu u_s(p) \Bigr]
 &= p^\lambda (d-4) \bigl( s^\rho g_{\alpha\beta} - s_\beta g^{\rho}_\alpha - s_\alpha g^{\rho}_\beta \Bigr)
\,.\end{align}
Here, we use naive dimensional regularization with anticommuting $\gamma^5$ and $p \cdot s = 0$, and furthermore that $\alpha$, $\beta$ and $\rho$ are transverse indices, while $\lambda$ is not.
\Eq{pretzelosity_trace} implies that  pretzelosity formally contributes at NLO 
but vanishes in dimensional regularization for $d=4$, so that the $\overline{\rm MS}$-renormalized pretzelosity quasi-TMD (and TMD) is zero at this order.
In fact, for matching using massless quark states, \refcites{Gutierrez-Reyes:2017glx,Gutierrez-Reyes:2018iod} observe that the pretzelosity matrix elements vanish through two loops, and \refcite{Chai:2018mwx} argues that this holds to all orders in perturbation theory.  
Recall that the requirement to carry out a valid matching calculation is that the chosen states have overlap with the operators being considered, which is not achieved with the above choices for pretzelosity.  Interestingly, we can consider matching using massless quark states with $d\ne 4$. In this case the $(d-4)$ prefactor occurs for both the quasi-TMD and TMD and cancels in their ratio, so we find
\begin{align}
 \tilde h_{1T}^\perp(x, b_T, \mu, P^z)
  = h_{1T}^\perp\bigl(x, b_T, \mu, \zeta=(2x P^z)^2 \bigr)
   \times \bigl[1 + \cO(\as)]
\,.\end{align}
In this matching relation the Wilson coefficient is only obtained at tree level, since the overlap matrix elements themselves start at one-loop.
To obtain the one-loop matching kernel for the quasi-pretzelosity in this fashion would require a two-loop calculation, which is beyond our goals here. However, the line of reasoning in \sec{allordermatch} suggests that $C_{\rm ns}^{h_{1T}^\perp}(\mu,xP^z)$ will also be universal with the same value as in \eq{Cnsall}.  A calculation of the ratio $\tilde h_{1T}^\perp / \tilde f_1$ would determine the size of the pretzelosity distribution relative to the unpolarized TMDPDF, which is an interesting target for lattice QCD.

\subsection{Generalization of matching to all orders}
\label{sec:allordermatch}

It is interesting to ask whether the observed spin independence of the one-loop matching kernels obtained in \eq{Cnsall} will continue to higher loop orders. In this section we outline an argument that this universality will continue to hold to all orders, without providing sufficient detail to call it a complete proof.

In the one-loop matching calculation we observed that it was the graphs associated to attachments to the Wilson lines that led to the mismatch between quasi-TMD and TMD contributions, and hence to a nonzero result for the matching coefficient. At leading order in power counting, these Wilson line graphs could not modify the spin structure of the initial operator. This is reminiscent of the spin universality of Soft Collinear Effective Theory (SCET) diagrams for heavy-to-light and back-to-back light-to-light currents~\cite{Bauer:2000yr,Bauer:2002nz,SCETlecture}, which arise from the spin universality of Wilson line and self-energy diagrams. 

Consider the generic beam function and quasi beam function correlators using quark fields with open Dirac spin indices $i$ and $j$,
\begin{align}
 B_{ij,q/h_S}(b^+,\bt,a,\tau,P^-) 
&=  
 \Bigl< h(P) \Bigr|  \Bigl[ \bar q_i(b^\mu) W_{\sqsubset}(b^\mu,0)
 q_j(0) \Bigr]_{\tau,a} \Bigl| h(P) \Bigr>
\,, \nn\\[5pt]
 \tilde B_{ij,q/h_S}(b^z,\bt,a,L,P^z) 
&= 
 \Bigl< h(P) \Bigr|  \Bigl[ \bar q_i(b^\mu) \widetilde W_{\sqsubset}(b^\mu,0;L)
 q_j(0) \Bigr]_{a} \Bigl| h(P) \Bigr>
 \,.
\end{align}
The paths for the Wilson lines are shown in \figs{wilsonlines}{qwilsonlines}, respectively. We notice that the vertices on the four corners of the Wilson lines are separated by long distances, either $\sim b_T$ or $\sim L$ for $\tilde B_{ij,q/h_S}$, or $\sim b_T$ or $\infty$ for $B_{ij,q/h_S}$.\footnote{Note that the situation for quasi-TMDs differs from that for matching spin-dependent quasi-PDFs~\cite{Xiong:2013bka,Izubuchi:2018srq,Alexandrou:2018eet,Liu:2018uuj,Liu:2018hxv}, since the quark fields in the quasi-PDF case are not separated by the same type of long distance scales.} These straight Wilson lines can be rewritten in terms of local operators by the now standard method of introducing scalar auxiliary fields $X_v$ for a line along the four-vector $v^\mu$~\cite{Korchemsky:1987wg}. For simplicity one can introduce three different auxiliary fields whose direction is along each straight Wilson line segment in \fig{qwilsonlines}, and another three auxiliary fields for the segments in \fig{wilsonlines}.  In this case both the quasi beam function and beam function calculations become matrix elements of a product of four local current operators whose spacetime arguments situate them at each of the corresponding four corners. We can refer to them as the current operators, such as $X_T^\dagger X_n$ and $X_n^\dagger q_i$, and quasi-current operators, such as $X_T^\dagger X_{\hat z}$ and $X_{\hat z}^\dagger q_i$. With this setup, two out of the four current operators (or quasi-current operators) have open spin indices ($i$ or $j$). 

When we carry out the matching in this setup we are matching one nonlocal product of operators onto another nonlocal product of operators; therefore one expects that the short-distance contribution is isolated to a matching coefficient $C$ at each of the currents when taking the boosted limit $X_{\hat z}^\dagger q_i \to C X_n^\dagger q_i$. This type of matching is like what occurs in SCET when matching a full theory current in boosted kinematics onto boosted $n$-collinear fields.  Due to the spin universality of the scalar auxiliary fields, no spin dependence will be introduced for the resulting short-distance matching coefficients $C$ at any order in $\alpha_s$. These arguments clearly must be extended to account for rapidity divergences that appear in the light-like limit, as well as non-trivial quasi soft and soft functions, but these are spin-independent and hence will not modify the spin universality of the matching coefficient. These rapidity factors and soft functions cancel out in ratios.

In the situation with infinite staple lengths, an analysis of this type that uses auxiliary fields was carried out recently in Ref.~\cite{Vladimirov:2020ofp}, from which we have drawn inspiration. The goal in that work was to derive the quasi-TMDPDF to TMDPDF matching relation.
Ref.~\cite{Vladimirov:2020ofp} uses SCET in a somewhat different fashion than what we envisioned above because they consider the analogy of the quasi-TMDPDF with a TMD hadronic tensor, and perform a match up which simultaneously yields $n$-collinear, $\bn$-collinear, and soft fields.  It is known that a quasi soft function is needed as part of the definition of the quasi-TMDPDF in order to properly carry out the quasi-TMDPDF to TMDPDF matching~\cite{Ji:2014hxa,Ji:2018hvs,Ebert:2019okf}, and it is so far not clear how the quasi soft function is treated by the analysis in~\refcite{Vladimirov:2020ofp}, which is also the case for our outline above.

\section{Applications}
\label{sec:applications}

We now discuss applications of our main finding in \eq{Cnsall}.
We find that the ratios of spin-dependent TMDs and the unpolarized TMD can be directly obtained from those of the quasi-TMDs, i.e.
\begin{align}
	   \frac{g_{1L}(x, b_T, \mu, \zeta)}{f_1(x, b_T, \mu, \zeta)} &
	= \frac{\tilde g_{1L}(x, b_T, \mu, P^z)}{\tilde f_1(x, b_T, \mu, P^z)}
   \,,\nn\\
	 \frac{h_1(x, b_T, \mu, \zeta)}{f_1(x, b_T, \mu, \zeta)} &
	 = \frac{\tilde h_1(x, b_T, \mu, P^z)}{\tilde f_1(x, b_T, \mu, P^z)}
  \,,\nn\\
	 \frac{h_{1T}^\perp(x, b_T, \mu, \zeta)}{f_1(x, b_T, \mu, \zeta)} &
	 = \frac{\tilde h_{1T}^\perp(x, b_T, \mu, P^z)}{\tilde f_1(x, b_T, \mu, P^z)}
    \,.
\end{align}
In these ratios the matching coefficients drop out along with the nonperturbative soft contributions in the function $g_q^S$ and the Collins-Soper evolution factor.  The anomalous dimensions for the $\mu$- and Collins-Soper evolutions are the same for the TMDs here, so the ratios on both the left- and the right-hand sides are only dependent on $x$ and $b_T$. These relations have power corrections that are suppressed by $1/(P^zb_T)$, so one can calculate the ratios on the r.h.s. in lattice QCD with different hadron momenta and interpolate to $P^z\to\infty$ to obtain the final result.

In addition, we can consider the ratios of the $x$-integrated TMDs which were studied in a different formalism based on exploiting Lorentz invariance in  Refs.~\cite{Musch:2010ka,Musch:2011er,Engelhardt:2015xja,Yoon:2016dyh,Yoon:2017qzo}.  According to \eq{qtmdpdf},
\begin{align} \label{eq:qtmdpdfx}
\int_{-1}^1 \!\df x\,\tilde f_{q/h_S}^{[\tilde\Gamma]}(x, \bt,\mu,P^z)
&= \tilde Z'_q(0,\mu,\tilde \mu) \tilde Z_{\rm uv}^q(0,\tilde \mu, a) \tilde\Delta_S^q(b_T, a, L)
   \tilde B_{q/h_S}^{[\tilde\Gamma]}(b^z=0, \bt, a, L, P^z)
\,,\end{align}
where $\tilde B_{q/h_S}^{[\tilde\Gamma]}(b^z=0, \bt, a, L, P^z)$ is the bare quasi beam function. The r.h.s.\ of \eq{qtmdpdfx} is finite, so the $x$-integration of the quasi-TMD is convergent. According to \eq{relation}, we  have
\begin{align}
 \frac{\int\!\df x\, \tilde F_{\ns/h_S}(x, b_T, \mu, P^z)}{\int\!\df x\,\tilde f_1(x, b_T, \mu, P^z)} &=  \frac{\tilde B_{\ns/h_S}(0, b_T, a,L, P^z)}{\tilde B_\ns(0, b_T, a, L, P^z)}\nn\\
 &=  \frac{\int\!\df x\,C_{\ns}\bigl(\mu, x P^z\bigr)\: F_{\ns/h_S}\big(x, b_T, \mu, \zeta=(2xP^z)^2\big)}{\int\!\df x\,C_{\ns}\bigl(\mu, x P^z\bigr)\: f_1\big(x, b_T, \mu, \zeta=(2xP^z)^2\big)}
  \,.
\end{align}
Thus we see that this ratio of $x$-integrated quasi-TMDPDF is not directly related to the ratio of $x$-integrated TMDs $F_{{\rm ns}/h_S}$ and $f_1$ in the formalism used here.

\section{Conclusion}
\label{sec:conclusion}

This paper constructs spin-dependent quasi-TMDPDFs that have a straightforward implementation in lattice calculations. We study the relationship between these spin-dependent quasi-TMDPDFs and their corresponding TMDPDFs at next-to-leading order.
The nonperturbative soft factor and Collins-Soper evolution cancel in the ratio of polarized and unpolarized quasi-TMDs, and in the ratio of the corresponding TMDs. This leads to a simple relationship between the quasi-TMD and TMD ratios involving only a perturbative matching kernel, see \eq{relation_ratio}. We calculate these kernels explicitly at one-loop in \sec{oneloop}.

At one-loop order in massless perturbation theory, we have access to the unpolarized ($f_1$), helicity ($g_{1L}$), and transversity ($h_1$) structures, which remarkably have the same perturbative kernel as the unpolarized case $C_{\ns}^{f_1} = C_\ns^{g_{1L}} = C_\ns^{h_1}$.
A key reason for this simple result is that all diagrams with nontrivial $x$ dependence must cancel (up to power corrections) between the perturbative results of quasi-TMDs and TMDs to give rise to an $x$-independent matching, as required by consistency~\cite{Ebert:2019okf}.
The remaining diagrams only contain the tree-level Dirac structure (which cancels out in suitable ratios) because the kinematic structures of all diagrams involving Wilson lines are independent of spin, up to power corrections.

An interesting feature of the pretzelosity is that its matrix element first appears at one-loop order, but vanishes in dimensional regularization when $d-4=-2\eps =0$. For $\epsilon\ne 0$ we can carry out matching between the  pretzelosity and quasi-pretzelosity, and they again satisfy the same relation as the aforementioned spin structures, yielding a simple matching relation at tree-level. The remaining spin structures, namely the Sivers function ($f_{1T}^\perp$), Boer-Mulders function ($h_1^\perp$) and worm-gear functions ($g_{1T}$, $h_{1L}^\perp$), vanish in massless perturbation theory  and are thus beyond the scope of this paper.

We also consider a definition of the quasi beam function with a staple-shaped Wilson line whose length remains constant under a Fourier transform.
This avoids the necessity of a $b^z$-dependent counterterm to cancel Wilson line self energies, and thus simplifies one aspect of lattice calculations of quasi-TMDPDF ratios.  We discuss implications of this definition for the determination of the Collins-Soper kernel in \app{sym}.

Finally, in \sec{allordermatch} we outline a procedure for generalizing the universality of the spin-dependent matching coefficients to all orders in perturbation theory.  We intend to carry out a more detailed analysis of this procedure in future work, as well as to obtain matching relations for the Sivers, Boer-Mulders, and worm-gear functions.

\begin{acknowledgments}
We thank Phiala Shanahan and Michael Wagman for useful discussions.
This work was supported by the U.S.\ Department of Energy, Office of Science,
Office of Nuclear Physics, from DE-SC0011090, DE-SC0012704
and within the framework of the TMD Topical Collaboration.
I.S.\ was also supported in part by the Simons Foundation through
the Investigator grant 327942.
M.E.\ was also supported by the Alexander von Humboldt Foundation
through a Feodor Lynen Research Fellowship.
S.T.S.\ was partially supported by the U.S. National Science Foundation through a Graduate Research Fellowship.
\end{acknowledgments}

\appendix

\section{Alternative construction of the quasi beam function}
\label{app:sym}

In \sec{def_qTMDPDF}, we mentioned that the definition of the quasi beam function in \eq{qbeam_b} has the nice feature that its renormalization factor is independent of $b^z$, which drops out in the ratios of quasi-TMDPDFs, simplifying the lattice calculations. In this appendix we contrast the definitions in \eq{qbeam_a} and \eq{qbeam_b} at a more technical level.

Take the Collins-Soper evolution kernel as an example. According to Refs.~\cite{Ebert:2018gzl,Ebert:2019tvc,Shanahan:2020zxr}, we can extract the nonperturbative Collins-Soper evolution kernel for TMDPDFs from the ratio of quasi-TMDPDFs at different hadron momenta for the definition in \eq{qbeam_a},
\begin{align} \label{eq:gamma_zeta}
&\gamma^q_\zeta(\mu, b_T) = \frac{1}{\ln(P^z_1/P^z_2)}
\\\nn&\times
\ln \frac{C_\ns(\mu,x P_2^z)\, \int\! \df b^z\, e^{ib^z xP_1^z}\, \tilde Z_q^\prime(b^z, \mu, \tilde\mu)\,
	\tilde Z_{\rm uv}^q(b^z,\tilde \mu, a) \, \cR_B(b_T,\tilde\mu,a,L) \,  \tilde B_\ns(b^z, \bt, a, P_1^z, L)}
{C_\ns(\mu,x P_1^z)\, \int\! \df b^z\, e^{ib^z xP_2^z}\, \tilde Z_q^\prime(b^z, \mu, \tilde\mu)\,
	\tilde Z_{\rm uv}^q(b^z,\tilde \mu, a) \, \cR_B(b_T,\tilde\mu,a,L) \, \tilde B_\ns(b^z, \bt, a, P_2^z, L)}
\,,\end{align}
where the quasi soft factor $\tilde \Delta_S^q$ cancels out, and the $b^z$-independent factor $\tilde R_B$ is constructed such that it exactly removes all divergences that would normally be canceled by $\tilde\Delta_S^q(b_T,a,L)$~\cite{Ebert:2019tvc}, \emph{i.e.}\ all power-law divergences not yet absorbed by $\tilde Z_{\rm uv}^q(b^z,\tilde\mu,a)$. In this way, we can take the $L\to\infty$ and $a\to0$ limits before forming the ratio.

If we instead use the alternative definition in \eq{qbeam_b}, both $\tilde Z_{\rm uv}^q(\tilde \mu, a)$ and $\tilde Z_q^\prime(\mu, \tilde\mu)$ also drop out in the ratios of the quasi-TMDPDFs. Therefore, if one loosens the requirement that the $L\to\infty$ and $a\to0$ limits be taken first, the Collins-Soper kernel can be obtained from the ratio of bare quasi beam functions on the lattice,
\begin{align} \label{eq:gamma_zeta_1}
\gamma^q_\zeta(\mu, b_T) &= \frac{1}{\ln(P^z_1/P^z_2)}\ln \left[\lim_{\substack{L\to\infty \\ a\to0}}\frac{C_\ns(\mu,x P_2^z)\, \int\! \df b^z\, e^{ib^z xP_1^z}\, \tilde B_{\ns}(b^z, \bt, a, P_1^z, L)}
{C_\ns(\mu,x P_1^z)\, \int\! \df b^z\, e^{ib^z xP_2^z}\, \tilde B_{\ns}(b^z, \bt, a, P_2^z, L)}\right]
\,.\end{align}
If one wants to take the $L\to\infty$ and $a\to0$ limits before forming the ratio, then $\tilde Z_{\rm uv}^q(\tilde \mu, a)$ and $\tilde Z_q^\prime(\mu, \tilde\mu)$ can be included and chosen as $\tilde Z_{\rm uv}^q(b^z=0,\tilde \mu, a)$ and $\tilde Z_q^\prime(b^z=0,\mu, \tilde\mu)$ which have already been studied in the RI/MOM scheme for the quasi beam function in \eq{qbeam}~\cite{Constantinou:2019vyb,Ebert:2019tvc,Shanahan:2019zcq}. Moreover, to exploit the $b^z$-independence, we might as well divide the bare quasi beam functions by $\tilde B_{\ns}(b^z=0, \bt, a, P^z, L)$, i.e.,
\begin{align} \label{eq:gamma_zeta_2}
&\gamma^q_\zeta(\mu, b_T) = \frac{1}{\ln(P^z_1/P^z_2)}\\
&\qquad\times 
\ln \frac{C_\ns(\mu,x P_2^z)\, \lim_{\substack{L\to\infty \\ a\to0}}\int\! \df b^z\, e^{ib^z xP_1^z}\, \tilde B_{\ns}(b^z, \bt, a, P_1^z, L)\big/\tilde B_{\ns}(0, \bt, a, P^z, L)}
{C_\ns(\mu,x P_1^z)\, \lim_{\substack{L\to\infty \\ a\to0}}\int\! \df b^z\, e^{ib^z xP_2^z}\, \tilde B_{\ns}(b^z, \bt, a, P_2^z, L)\big/\tilde B_{\ns}(0, \bt, a, P^z, L)}
\,.\nn\end{align}
This has the advantage that the errors at $b^z=0$ and $b^z\neq0$ are correlated and can be reduced by the division. Here $P^z$ can take on any value, and to reduce the errors one can simply choose $P^z=0$.

As a word of caution, we note that the discussion above did not take into account the mixing of $\tilde B_{\ns/h_S}^{\tilde\Gamma}$ with other Dirac structures on the lattice~\cite{Constantinou:2019vyb,Shanahan:2019zcq,Green:2020xco}. According to the RI/MOM analysis in Ref.~\cite{Shanahan:2019zcq}, mixing effects can become considerable for certain lattice ensembles which are $b^z$ dependent. Therefore, it may still be inevitable that one must carry out the procedure of diagonalizing the renormalization matrix and converting to the $\MS$ scheme as in Refs.~\cite{Shanahan:2019zcq,Shanahan:2020zxr} to extract the Collins-Soper kernel or other types of ratios. Nevertheless, if the lattice parameters could be fine tuned to make the operator mixing effect less important than other systematic errors, then we can exploit the above simplification.


\section{Independent calculation of the spin-dependent quasi-TMDPDFs}
\label{app:SDTMDPDF}

In this section we perform an independent calculation of the one-loop quasi-TMDPDFs by postponing the approximations used in \eqs{sail_dirac_2}{sail_dirac_3} after the loop integration. Our strategy is to express the spin-dependent quasi-TMDPDFs as a linear combination of the unpolarized one and novel structures, the latter of which turn out to be two terms.

\subsection{Vertex diagram}
\label{app:vertex}

In this subsection, we explicitly evaluate the vertex diagram for all relevant spin structures.
\Eq{qtmd_a_2} gives the exact result for the vertex diagram as
\begin{align}
 \tilde q_A^{\tilde\Gamma\,(1)}(b, p) &
 = \Tr\Bigl[\bar u_s(p) \gamma^\mu \gamma_\alpha \frac{\tilde\Gamma}{2} \gamma_\beta \gamma_\mu u_s(p) \Bigr]
   \frac{\as C_F}{4\pi} \frac{\Gamma(-\eps)}{e^{\eps\gamma_E}} e^{\eps L_b}
   \nn\\&\quad\times
   \biggl(\frac{g_\perp^{\alpha\beta}}{2} + \eps \frac{b^\alpha b^\beta}{b^2} \biggr)
   \biggl(\frac{-b^2}{b_T^2}\biggr)^{\eps} \frac{1 + \img \pb - e^{\img \pb}}{(\pb)^2}
\,.\end{align}
Using the completeness relation for polarized spinors given in \eq{completeness_massless}, the Dirac traces for the different choices in \eq{qGamma} yield
\begin{alignat}{3} \label{eq:vertex_traces}
 &\tilde\Gamma = \gamma^\lambda: \quad
 &&\Tr\biggl[\bar u_s(p) \gamma^\mu \gamma_\alpha \frac{\tilde\Gamma}{2} \gamma_\beta \gamma_\mu u_s(p) \biggr]
 &&= (d-2)\bigl(p^\lambda g_{\alpha\beta} - p_\beta g_\alpha^\lambda -p_\alpha g_\beta^\lambda \bigr)
\,,\nn\\
 &\tilde\Gamma = \gamma^\lambda\gamma_5: \quad
 &&\Tr\biggl[\bar u_s(p) \gamma^\mu \gamma_\alpha \frac{\tilde\Gamma}{2} \gamma_\beta \gamma_\mu u_s(p) \biggr]
 &&= \Lambda (d-2)\bigl(p^\lambda g_{\alpha\beta} - p_\beta g_\alpha^\lambda -p_\alpha g_\beta^\lambda \bigr)
\,,\nn\\
 &\tilde\Gamma = \img \sigma^{\sigma \lambda} \gamma_5: \quad
 &&\Tr\biggl[\bar u_s(p) \gamma^\mu \gamma_\alpha \frac{\tilde\Gamma}{2} \gamma_\beta \gamma_\mu u_s(p) \biggr]
 &&=(d-4)\Big[ s^\sigma_\perp \bigl(p^\lambda g_{\alpha\beta} - p_\beta g_\alpha^\lambda -p_\alpha g_\beta^\lambda\bigr)
   \nn\\&&&&&\hspace{1.8cm}
   -p^\lambda \bigl(s_{\perp\alpha} g^\sigma_\beta + s_{\perp\beta} g^\sigma_\alpha \bigr)\Bigr]
\,.\end{alignat}
The first two traces are equivalent; thus, the unpolarized and helicity quasi-TMDs are identical and are given by\begin{align} \label{eq:bla1}
 \tilde q_A^{\gamma^\lambda\,(1)}(b, p) &
 = \tilde q_A^{\gamma^\lambda\gamma_5 \,(1)}(b, p)
 \nn\\&
 = \frac{\as C_F}{4\pi} \frac{\Gamma(-\eps)}{e^{\eps\gamma_E}} e^{\eps L_b}
   (d-2) \left(p^\lambda - 2 \eps \frac{ \pb b^\lambda}{b^2}\right)
   \biggl(\frac{-b^2}{b_T^2}\biggr)^{\eps} \frac{1 + \img \pb - e^{\img \pb}}{(\pb)^2}
\,.\end{align}
For $\lambda=t$, this reduces to the case already calculated in \refcite{Ebert:2019okf}.
The difference between $\lambda=t$ and $\lambda=z$ is given by the second term in \eq{bla1}, which is finite and can be evaluated after setting $\eps=0$.
Applying the Fourier transform, it evaluates to
\begin{align} \label{eq:bla2}
 & \tilde q_A^{\gamma^t\,(1)}(x,\bt) - \tilde q_A^{\gamma^z\,(1)}(x,\bt)
 \nn\\&
 = \frac{\as C_F}{\pi} \frac{1}{p^z} \int\frac{\df b^z}{2\pi} e^{\img x b^z p^z}
   \frac{1 - \img p^z b^z - e^{-\img p^z b^z}}{b_T^2 + b_z^2}
 \nn\\&
 = \frac{\as C_F}{2 \pi} \frac{1}{b_T p^z} \Bigl\{
   [1 + b_T p^z {\rm sgn}(x)] e^{-|x| b_T p^z} - e^{-|1-x| b_T p^z} \Bigr\}
\,.\end{align}
This term is exponentially suppressed for large $b_T p^z \gg 1$ and thus can be neglected.

The last Dirac structure in \eq{vertex_traces} consists of a piece proportional to the Dirac structure of the unpolarized TMD and a genuinely new structure, which we can evaluate for $\eps\to 0$ using \eq{int_vertex},
\begin{align}
 \tilde q_A^{\img \sigma^{\sigma \lambda} \gamma_5,(1)}(x,\bt) &
 = \frac{d-4}{d-2} s^\sigma_\perp \tilde q_A^{\gamma^\lambda\,(1)}(x,\bt)
   - 2 s_\perp^\sigma \frac{\as C_F}{4\pi}(1-x) \theta(x) \theta(1-x)
 = 0
\,.\end{align}

\subsection{Sail diagram}
\label{app:sail}

The sail diagram is given in \eq{qtmd_b_1} as
\begin{align} \label{eq:sail_app}
 \tilde q_B^{\tilde\Gamma\,(1)}(b) &
 = -\frac{\as C_F}{2\pi} \Tr\bigl[ \bar u_s(p) \tilde\Gamma \gamma^\rho \gamma^\mu u_s(p) \bigr]
  \frac{\mu_0^{2\eps}}{(2\pi)^{d-2}} \int\!\df^d k \int_0^1 \df s \,
  \frac{\gamma'(s)_\mu k_\rho}{k^2 (p-k)^2} \, e^{\img p \cdot b -\img (p-k) \cdot \gamma(s)}
 \nn\\&\quad
 - \frac{\as C_F}{2\pi} \Tr\bigl[ \bar u_s(p) \gamma^\mu \gamma^\rho \tilde\Gamma u_s(p) \bigr]
  \frac{\mu_0^{2\eps}}{(2\pi)^{d-2}} \int\!\df^d k \int_0^1 \df s \,
  \frac{\gamma'(s)_\mu k_\rho}{k^2 (p-k)^2} \, e^{\img k \cdot b + \img (p-k)\cdot\gamma(s)}
\,.\end{align}
Evaluating the Dirac trace for the different Dirac structures given in \eq{qGamma}, we have
\begin{alignat}{3} \label{eq:sail_traces}
 &\tilde\Gamma = \gamma^\lambda: \quad
 &&\Tr\bigl[ \bar u_s(p) \tilde\Gamma \gamma^\rho \gamma^\mu u_s(p) \bigr]
 &&= p^\mu g^{\rho\lambda} -p^\rho g^{\lambda\mu} +p^\lambda g^{\rho\mu}
\,,\\\nn
 &\tilde\Gamma = \gamma^\lambda\gamma_5: \quad
 &&\Tr\bigl[ \bar u_s(p) \tilde\Gamma \gamma^\rho \gamma^\mu u_s(p) \bigr]
 &&= \Lambda \bigl(p^\mu g^{\rho\lambda} -p^\rho g^{\lambda\mu} +p^\lambda g^{\rho\mu}\bigr)
\,,\\\nn
 &\tilde\Gamma = \img \sigma^{\sigma \lambda} \gamma_5: \quad
 &&\Tr\bigl[ \bar u_s(p) \tilde\Gamma \gamma^\rho \gamma^\mu u_s(p) \bigr]
 &&= s_\perp^\sigma \bigl( p^\mu g^{\rho\lambda} -p^\rho g^{\lambda\mu} +p^\lambda g^{\rho\mu} \bigr)
   - p^\lambda \bigl(s^{\rho}_\perp g^{\mu\sigma } + s^{\mu }_\perp g^{\rho\sigma } \bigr)
\,.\end{alignat}
We obtain the same results for the trace in the second line of \eq{sail_app}.
The first two Dirac structures in \eq{sail_traces} yield the same traces, and thus the vertex diagram results for the unpolarized and helicity structures are identical.
It is also easy to see from the structure of the integrand that $\lambda=t$ and $\lambda=z$ yield identical results, as also discussed in \refcite{Ebert:2019okf}.

Using \eq{sail_traces}, we can relate the transversity to the unpolarized TMD by
\begin{align} \label{eq:sail_h_diff}
 \left[s_\perp^\sigma \tilde h_1^{\lambda(1)}(x,\bt)
 + \biggl(\frac{g_\perp^{\sigma\rho}}{2} - \frac{b_\perp^\sigma b_\perp^\rho}{b_\perp^2}\biggr) s_{\perp\,\rho} \tilde h_{1T}^{\lambda\perp(1)}(x,\bt) \right]_{\rm s}
 = s_\perp^\sigma \tilde f_{1,\rm s}^{\lambda(1)}(x,\bt)
 + \Delta\tilde h_{\rm s}^{\lambda\sigma}(x,\bt)
\,,\end{align}
where prior to the Fourier transform
\begin{align} \label{eq:sail_h_diff_2}
 \Delta\tilde h_{\rm s}^{\lambda\sigma}(p,b) &
 = \frac{\as C_F}{2\pi}  p^\lambda \bigl(s^{\rho}_\perp g^{\mu\sigma } + s^{\mu }_\perp g^{\rho\sigma } \bigr)
  \int_0^1 \df s \, \gamma'(s)_\mu
  \nn\\*&\quad\times
  \frac{\mu_0^{2\eps}}{(2\pi)^{d-2}}  \int\!\df^d k  \frac{ k_\rho}{k^2 (p-k)^2} \, \Bigl[ e^{\img p \cdot [b - \gamma(s)]} e^{\img k \cdot \gamma(s)} + e^{\img k \cdot [b-\gamma(s)]} e^{\img p\cdot\gamma(s)} \Bigr]
\,.\end{align}
The two integrals over $\df^d k$ can be evaluated using
\begin{align} \label{eq:blubb}
 \frac{\mu_0^{2\eps}}{(2\pi)^{d-2}} \int\!\df^dk \frac{k_\rho e^{\img k \cdot x}}{k^2 (p-k)^2} &
 = \frac{\img}{4} \frac{\Gamma(-\eps)}{e^{\eps \gamma_E}} \left(\frac{- \mu^2 x^2}{b_0^2}\right)^\eps
   \left[ \frac{2 \eps x_\rho}{x^2} \frac{1 - e^{\img p \cdot x}}{p \cdot x}
   + p_\rho \frac{e^{\img p \cdot x}(1-\img p \cdot x) - 1}{(p \cdot x)^2} \right]
\,.\end{align}
Since the spin structure in \eq{sail_h_diff_2} is purely transverse, we have $p_\rho = 0$, and thus only the first structure in \eq{blubb} contributes.
We then take $\eps\to0$, which yields
\begin{align} \label{eq:sail_h_diff_3}
 \Delta\tilde h_{\rm s}^{\lambda\sigma}(p,b) &
 = -\frac{\img}{2} \frac{\as C_F}{2\pi} p^\lambda \bigl(s^{\rho}_\perp g^{\mu\sigma } + s^{\mu }_\perp g^{\rho\sigma } \bigr)
  e^{\img p \cdot b} \int_0^1 \df s \, \gamma'(s)_\mu
  \nn\\&\quad\times
  \biggl\{ \frac{\gamma_\rho(s)}{\gamma^2(s)} \frac{e^{-\img p \cdot \gamma(s)} - 1}{p \cdot \gamma(s)}
  + \frac{[b-\gamma(s)]_\rho}{[b-\gamma(s)]^2} \frac{e^{\img p\cdot[\gamma(s)-b]}  - 1}{p \cdot [b-\gamma(s)]} \biggr\}
\,.\end{align}
Since $\gamma'_\mu$ contracts with a transverse vector, only the perpendicular Wilson line segment contributes; its parametrization takes the form
\begin{align}
 \gamma(s)^\mu = (0, s \bt, L)
\,.\end{align}
Again using that both $\rho$ and $\sigma$ are transverse indices, we obtain
\begin{align} \label{eq:sail_h_diff_4}
 \Delta\tilde h_{\rm s}^{\lambda\sigma}(p,b) &
 = -\img \frac{\as C_F}{2\pi} p^\lambda (s_\perp{\cdot}b_\perp) b_\perp^\sigma
  e^{\img p \cdot b} \int_0^1 \df s \,
  \biggl[ \frac{s}{s^2 b_\perp^2 - L^2} \frac{1 - e^{\img p^z L}}{p^z L} + (L \to b^z - L) \biggr]
\nn\\&
 = -\frac{\img}{2} \frac{\as C_F}{2\pi} p^\lambda \frac{(s_\perp{\cdot}b_\perp) b_\perp^\sigma}{b_\perp^2}
  e^{\img p \cdot b} \biggl[ \ln\Bigl(1 - \frac{b_\perp^2}{L^2}\Bigr)  \frac{1 - e^{\img p^z L}}{p^z L} + (L \to b^z - L) \biggr]
\,.\end{align}
Even prior to evaluating the Fourier transform, it is clear that this expression vanishes in the limit of large $p^z L , L/b_T \gg 1$.
Thus, in conclusion, we have that the sail diagram yields
\begin{align}
 \tilde f_{1,\rm s}^{\lambda (1)}(x,b_T) = \tilde g_{1L,\rm s}^{\lambda(1)}(x,b_T) = \tilde h_{1,\rm s}^{\lambda(1)}(x,b_T)
\,,\end{align}
while all other structures vanish.

\bibliographystyle{JHEP}
\bibliography{literature}

\providecommand{\href}[2]{#2}\begingroup\raggedright\begin{thebibliography}{10}

\bibitem{Gautheron:2010wva}
{\scshape COMPASS} collaboration, F.~Gautheron et~al., \emph{{COMPASS-II
  Proposal}}, .

\bibitem{Dudek:2012vr}
J.~Dudek et~al., \emph{{Physics Opportunities with the 12 GeV Upgrade at
  Jefferson Lab}}, \href{https://doi.org/10.1140/epja/i2012-12187-1}{\emph{Eur.
  Phys. J. A} {\bfseries 48} (2012) 187}
  [\href{https://arxiv.org/abs/1208.1244}{{\ttfamily 1208.1244}}].

\bibitem{Aschenauer:2015eha}
E.-C. Aschenauer et~al., \emph{{The RHIC SPIN Program: Achievements and Future
  Opportunities}},  \href{https://arxiv.org/abs/1501.01220}{{\ttfamily
  1501.01220}}.

\bibitem{Accardi:2012qut}
A.~Accardi et~al., \emph{{Electron Ion Collider: The Next QCD Frontier}},
  {\emph{Eur. Phys. J.} {\bfseries A52} (2016) 268}
  [\href{https://arxiv.org/abs/1212.1701}{{\ttfamily 1212.1701}}].

\bibitem{Bacchetta:2017gcc}
A.~Bacchetta, F.~Delcarro, C.~Pisano, M.~Radici and A.~Signori,
  \emph{{Extraction of partonic transverse momentum distributions from
  semi-inclusive deep-inelastic scattering, Drell-Yan and Z-boson production}},
  \href{https://doi.org/10.1007/JHEP06(2017)081}{\emph{JHEP} {\bfseries 06}
  (2017) 081} [\href{https://arxiv.org/abs/1703.10157}{{\ttfamily
  1703.10157}}].

\bibitem{Scimemi:2017etj}
I.~Scimemi and A.~Vladimirov, \emph{{Analysis of vector boson production within
  TMD factorization}}, {\emph{Eur. Phys. J.} {\bfseries C78} (2018) 89}
  [\href{https://arxiv.org/abs/1706.01473}{{\ttfamily 1706.01473}}].

\bibitem{Bertone:2019nxa}
V.~Bertone, I.~Scimemi and A.~Vladimirov, \emph{{Extraction of unpolarized
  quark transverse momentum dependent parton distributions from
  Drell-Yan/Z-boson production}},
  \href{https://doi.org/10.1007/JHEP06(2019)028}{\emph{JHEP} {\bfseries 06}
  (2019) 028} [\href{https://arxiv.org/abs/1902.08474}{{\ttfamily
  1902.08474}}].

\bibitem{Scimemi:2019cmh}
I.~Scimemi and A.~Vladimirov, \emph{{Non-perturbative structure of
  semi-inclusive deep-inelastic and Drell-Yan scattering at small transverse
  momentum}},  \href{https://arxiv.org/abs/1912.06532}{{\ttfamily 1912.06532}}.

\bibitem{Bacchetta:2019sam}
A.~Bacchetta, V.~Bertone, C.~Bissolotti, G.~Bozzi, F.~Delcarro, F.~Piacenza
  et~al., \emph{{Transverse-momentum-dependent parton distributions up to
  N$^3$LL from Drell-Yan data}},
  \href{https://arxiv.org/abs/1912.07550}{{\ttfamily 1912.07550}}.

\bibitem{1793441}
A.~Bacchetta, F.~Delcarro, C.~Pisano and M.~Radici, \emph{{The
  three-dimensional distribution of quarks in momentum space}},
  \href{https://arxiv.org/abs/2004.14278}{{\ttfamily 2004.14278}}.

\bibitem{Ji:2013dva}
X.~Ji, \emph{{Parton Physics on a Euclidean Lattice}}, {\emph{Phys. Rev. Lett.}
  {\bfseries 110} (2013) 262002}
  [\href{https://arxiv.org/abs/1305.1539}{{\ttfamily 1305.1539}}].

\bibitem{Ji:2014gla}
X.~Ji, \emph{{Parton Physics from Large-Momentum Effective Field Theory}},
  {\emph{Sci. China Phys. Mech. Astron.} {\bfseries 57} (2014) 1407}
  [\href{https://arxiv.org/abs/1404.6680}{{\ttfamily 1404.6680}}].

\bibitem{Ji:2020ect}
X.~Ji, Y.-S. Liu, Y.~Liu, J.-H. Zhang and Y.~Zhao, \emph{{Large-Momentum
  Effective Theory}},  \href{https://arxiv.org/abs/2004.03543}{{\ttfamily
  2004.03543}}.

\bibitem{Ji:2014hxa}
X.~Ji, P.~Sun, X.~Xiong and F.~Yuan, \emph{{Soft factor subtraction and
  transverse momentum dependent parton distributions on the lattice}},
  {\emph{Phys. Rev.} {\bfseries D91} (2015) 074009}
  [\href{https://arxiv.org/abs/1405.7640}{{\ttfamily 1405.7640}}].

\bibitem{Ji:2018hvs}
X.~Ji, L.-C. Jin, F.~Yuan, J.-H. Zhang and Y.~Zhao, \emph{{Transverse Momentum
  Dependent Quasi-Parton-Distributions}}, {\emph{Phys. Rev.} {\bfseries D99}
  (2019) 114006} [\href{https://arxiv.org/abs/1801.05930}{{\ttfamily
  1801.05930}}].

\bibitem{Ebert:2018gzl}
M.~A. Ebert, I.~W. Stewart and Y.~Zhao, \emph{{Determining the Nonperturbative
  Collins-Soper Kernel From Lattice QCD}}, {\emph{Phys. Rev.} {\bfseries D99}
  (2019) 034505} [\href{https://arxiv.org/abs/1811.00026}{{\ttfamily
  1811.00026}}].

\bibitem{Ebert:2019okf}
M.~A. Ebert, I.~W. Stewart and Y.~Zhao, \emph{{Towards Quasi-Transverse
  Momentum Dependent PDFs Computable on the Lattice}},
  \href{https://doi.org/10.1007/JHEP09(2019)037}{\emph{JHEP} {\bfseries 09}
  (2019) 037} [\href{https://arxiv.org/abs/1901.03685}{{\ttfamily
  1901.03685}}].

\bibitem{Ebert:2019tvc}
M.~A. Ebert, I.~W. Stewart and Y.~Zhao, \emph{{Renormalization and Matching for
  the Collins-Soper Kernel from Lattice QCD}},
  \href{https://doi.org/10.1007/JHEP03(2020)099}{\emph{JHEP} {\bfseries 03}
  (2020) 099} [\href{https://arxiv.org/abs/1910.08569}{{\ttfamily
  1910.08569}}].

\bibitem{Ji:2019sxk}
X.~Ji, Y.~Liu and Y.-S. Liu, \emph{{QCD Soft Function from Large-Momentum
  Effective Theory on Lattice}},
  \href{https://arxiv.org/abs/1910.11415}{{\ttfamily 1910.11415}}.

\bibitem{Ji:2019ewn}
X.~Ji, Y.~Liu and Y.-S. Liu, \emph{{Transverse-Momentum-Dependent PDFs from
  Large-Momentum Effective Theory}},
  \href{https://arxiv.org/abs/1911.03840}{{\ttfamily 1911.03840}}.

\bibitem{Vladimirov:2020ofp}
A.~A. Vladimirov and A.~Sch{\"a}fer, \emph{{Transverse momentum dependent
  factorization for lattice observables}},
  \href{https://arxiv.org/abs/2002.07527}{{\ttfamily 2002.07527}}.

\bibitem{Musch:2010ka}
B.~U. Musch, P.~H{\"a}gler, J.~W. Negele and A.~Sch{\"a}fer, \emph{{Exploring
  quark transverse momentum distributions with lattice QCD}}, {\emph{Phys.
  Rev.} {\bfseries D83} (2011) 094507}
  [\href{https://arxiv.org/abs/1011.1213}{{\ttfamily 1011.1213}}].

\bibitem{Musch:2011er}
B.~U. Musch, P.~H{\"a}gler, M.~Engelhardt, J.~W. Negele and A.~Sch{\"a}fer,
  \emph{{Sivers and Boer-Mulders observables from lattice QCD}}, {\emph{Phys.
  Rev.} {\bfseries D85} (2012) 094510}
  [\href{https://arxiv.org/abs/1111.4249}{{\ttfamily 1111.4249}}].

\bibitem{Engelhardt:2015xja}
M.~Engelhardt, P.~H{\"a}gler, B.~Musch, J.~Negele and A.~Sch{\"a}fer,
  \emph{{Lattice QCD study of the Boer-Mulders effect in a pion}}, {\emph{Phys.
  Rev.} {\bfseries D93} (2016) 054501}
  [\href{https://arxiv.org/abs/1506.07826}{{\ttfamily 1506.07826}}].

\bibitem{Yoon:2016dyh}
B.~Yoon, T.~Bhattacharya, M.~Engelhardt, J.~Green, R.~Gupta, P.~H{\"a}gler
  et~al., \emph{{Lattice QCD calculations of nucleon transverse
  momentum-dependent parton distributions using clover and domain wall
  fermions}},  in \emph{{Proceedings, 33rd International Symposium on Lattice
  Field Theory (Lattice 2015): Kobe, Japan, July 14-18, 2015}}, SISSA, SISSA,
  2015, \href{https://arxiv.org/abs/1601.05717}{{\ttfamily 1601.05717}}.

\bibitem{Yoon:2017qzo}
B.~Yoon, M.~Engelhardt, R.~Gupta, T.~Bhattacharya, J.~R. Green, B.~U. Musch
  et~al., \emph{{Nucleon Transverse Momentum-dependent Parton Distributions in
  Lattice QCD: Renormalization Patterns and Discretization Effects}},
  {\emph{Phys. Rev.} {\bfseries D96} (2017) 094508}
  [\href{https://arxiv.org/abs/1706.03406}{{\ttfamily 1706.03406}}].

\bibitem{Shanahan:2019zcq}
P.~Shanahan, M.~L. Wagman and Y.~Zhao, \emph{{Nonperturbative renormalization
  of staple-shaped Wilson line operators in lattice QCD}},
  \href{https://doi.org/10.1103/PhysRevD.101.074505}{\emph{Phys. Rev.}
  {\bfseries D101} (2020) 074505}
  [\href{https://arxiv.org/abs/1911.00800}{{\ttfamily 1911.00800}}].

\bibitem{Shanahan:2020zxr}
P.~Shanahan, M.~Wagman and Y.~Zhao, \emph{{Collins-Soper Kernel for TMD
  Evolution from Lattice QCD}},
  \href{https://arxiv.org/abs/2003.06063}{{\ttfamily 2003.06063}}.

\bibitem{Collins:1981va}
J.~C. Collins and D.~E. Soper, \emph{{Back-To-Back Jets: Fourier Transform from
  B to K-Transverse}}, {\emph{Nucl. Phys.} {\bfseries B197} (1982) 446}.

\bibitem{Collins:1981uk}
J.~C. Collins and D.~E. Soper, \emph{{Back-To-Back Jets in QCD}}, {\emph{Nucl.
  Phys.} {\bfseries B193} (1981) 381}.

\bibitem{Manohar:2006nz}
A.~V. Manohar and I.~W. Stewart, \emph{{The Zero-Bin and Mode Factorization in
  Quantum Field Theory}}, {\emph{Phys. Rev.} {\bfseries D76} (2007) 074002}
  [\href{https://arxiv.org/abs/hep-ph/0605001}{{\ttfamily hep-ph/0605001}}].

\bibitem{Collins:1350496}
J.~Collins, \emph{{Foundations of perturbative QCD}}, Cambridge monographs on
  particle physics, nuclear physics, and cosmology. Cambridge Univ. Press, New
  York, NY, 2011.

\bibitem{Beneke:2003pa}
M.~Beneke and T.~Feldmann, \emph{{Factorization of heavy to light form-factors
  in soft collinear effective theory}}, {\emph{Nucl. Phys.} {\bfseries B685}
  (2004) 249} [\href{https://arxiv.org/abs/hep-ph/0311335}{{\ttfamily
  hep-ph/0311335}}].

\bibitem{Chiu:2007yn}
J.-y. Chiu, F.~Golf, R.~Kelley and A.~V. Manohar, \emph{{Electroweak Sudakov
  corrections using effective field theory}}, {\emph{Phys. Rev. Lett.}
  {\bfseries 100} (2008) 021802}
  [\href{https://arxiv.org/abs/0709.2377}{{\ttfamily 0709.2377}}].

\bibitem{Becher:2011dz}
T.~Becher and G.~Bell, \emph{{Analytic Regularization in Soft-Collinear
  Effective Theory}}, {\emph{Phys. Lett.} {\bfseries B713} (2012) 41}
  [\href{https://arxiv.org/abs/1112.3907}{{\ttfamily 1112.3907}}].

\bibitem{Chiu:2011qc}
J.-y. Chiu, A.~Jain, D.~Neill and I.~Z. Rothstein, \emph{{The Rapidity
  Renormalization Group}}, {\emph{Phys. Rev. Lett.} {\bfseries 108} (2012)
  151601} [\href{https://arxiv.org/abs/1104.0881}{{\ttfamily 1104.0881}}].

\bibitem{Chiu:2012ir}
J.-Y. Chiu, A.~Jain, D.~Neill and I.~Z. Rothstein, \emph{{A Formalism for the
  Systematic Treatment of Rapidity Logarithms in Quantum Field Theory}},
  {\emph{JHEP} {\bfseries 05} (2012) 084}
  [\href{https://arxiv.org/abs/1202.0814}{{\ttfamily 1202.0814}}].

\bibitem{Chiu:2009yx}
J.-y. Chiu, A.~Fuhrer, A.~H. Hoang, R.~Kelley and A.~V. Manohar,
  \emph{{Soft-Collinear Factorization and Zero-Bin Subtractions}}, {\emph{Phys.
  Rev.} {\bfseries D79} (2009) 053007}
  [\href{https://arxiv.org/abs/0901.1332}{{\ttfamily 0901.1332}}].

\bibitem{GarciaEchevarria:2011rb}
M.~G. Echevarria, A.~Idilbi and I.~Scimemi, \emph{{Factorization Theorem For
  Drell-Yan At Low $q_T$ And Transverse Momentum Distributions
  On-The-Light-Cone}}, {\emph{JHEP} {\bfseries 07} (2012) 002}
  [\href{https://arxiv.org/abs/1111.4996}{{\ttfamily 1111.4996}}].

\bibitem{Li:2016axz}
Y.~Li, D.~Neill and H.~X. Zhu, \emph{{An Exponential Regulator for Rapidity
  Divergences}},  \href{https://arxiv.org/abs/1604.00392}{{\ttfamily
  1604.00392}}.

\bibitem{Ji:2004wu}
X.-d. Ji, J.-p. Ma and F.~Yuan, \emph{{QCD factorization for semi-inclusive
  deep-inelastic scattering at low transverse momentum}}, {\emph{Phys. Rev.}
  {\bfseries D71} (2005) 034005}
  [\href{https://arxiv.org/abs/hep-ph/0404183}{{\ttfamily hep-ph/0404183}}].

\bibitem{Ji:2002aa}
X.-d. Ji and F.~Yuan, \emph{{Parton distributions in light cone gauge: Where
  are the final state interactions?}}, {\emph{Phys. Lett.} {\bfseries B543}
  (2002) 66} [\href{https://arxiv.org/abs/hep-ph/0206057}{{\ttfamily
  hep-ph/0206057}}].

\bibitem{Belitsky:2002sm}
A.~V. Belitsky, X.~Ji and F.~Yuan, \emph{{Final state interactions and gauge
  invariant parton distributions}}, {\emph{Nucl. Phys.} {\bfseries B656} (2003)
  165} [\href{https://arxiv.org/abs/hep-ph/0208038}{{\ttfamily
  hep-ph/0208038}}].

\bibitem{Idilbi:2010im}
A.~Idilbi and I.~Scimemi, \emph{{Singular and Regular Gauges in Soft Collinear
  Effective Theory: The Introduction of the New Wilson Line T}}, {\emph{Phys.
  Lett.} {\bfseries B695} (2011) 463}
  [\href{https://arxiv.org/abs/1009.2776}{{\ttfamily 1009.2776}}].

\bibitem{GarciaEchevarria:2011md}
M.~Garcia-Echevarria, A.~Idilbi and I.~Scimemi, \emph{{SCET, Light-Cone Gauge
  and the T-Wilson Lines}}, {\emph{Phys. Rev.} {\bfseries D84} (2011) 011502}
  [\href{https://arxiv.org/abs/1104.0686}{{\ttfamily 1104.0686}}].

\bibitem{Ralston:1979ys}
J.~P. Ralston and D.~E. Soper, \emph{{Production of Dimuons from High-Energy
  Polarized Proton Proton Collisions}},
  \href{https://doi.org/10.1016/0550-3213(79)90082-8}{\emph{Nucl. Phys.}
  {\bfseries B152} (1979) 109}.

\bibitem{Tangerman:1994eh}
R.~D. Tangerman and P.~J. Mulders, \emph{{Intrinsic transverse momentum and the
  polarized Drell-Yan process}},
  \href{https://doi.org/10.1103/PhysRevD.51.3357}{\emph{Phys. Rev.} {\bfseries
  D51} (1995) 3357} [\href{https://arxiv.org/abs/hep-ph/9403227}{{\ttfamily
  hep-ph/9403227}}].

\bibitem{Boer:1997nt}
D.~Boer and P.~J. Mulders, \emph{{Time reversal odd distribution functions in
  leptoproduction}}, {\emph{Phys. Rev.} {\bfseries D57} (1998) 5780}
  [\href{https://arxiv.org/abs/hep-ph/9711485}{{\ttfamily hep-ph/9711485}}].

\bibitem{Mulders:1995dh}
P.~J. Mulders and R.~D. Tangerman, \emph{{The Complete tree level result up to
  order 1/Q for polarized deep inelastic leptoproduction}}, {\emph{Nucl. Phys.}
  {\bfseries B461} (1996) 197}
  [\href{https://arxiv.org/abs/hep-ph/9510301}{{\ttfamily hep-ph/9510301}}].

\bibitem{Bacchetta:2004zf}
A.~Bacchetta, P.~J. Mulders and F.~Pijlman, \emph{{New observables in
  longitudinal single-spin asymmetries in semi-inclusive DIS}},
  \href{https://doi.org/10.1016/j.physletb.2004.06.052}{\emph{Phys. Lett.}
  {\bfseries B595} (2004) 309}
  [\href{https://arxiv.org/abs/hep-ph/0405154}{{\ttfamily hep-ph/0405154}}].

\bibitem{Goeke:2005hb}
K.~Goeke, A.~Metz and M.~Schlegel, \emph{{Parameterization of the quark-quark
  correlator of a spin-1/2 hadron}},
  \href{https://doi.org/10.1016/j.physletb.2005.05.037}{\emph{Phys. Lett.}
  {\bfseries B618} (2005) 90}
  [\href{https://arxiv.org/abs/hep-ph/0504130}{{\ttfamily hep-ph/0504130}}].

\bibitem{Bacchetta:2006tn}
A.~Bacchetta, M.~Diehl, K.~Goeke, A.~Metz, P.~J. Mulders and M.~Schlegel,
  \emph{{Semi-inclusive deep inelastic scattering at small transverse
  momentum}}, {\emph{JHEP} {\bfseries 02} (2007) 093}
  [\href{https://arxiv.org/abs/hep-ph/0611265}{{\ttfamily hep-ph/0611265}}].

\bibitem{Sivers:1989cc}
D.~W. Sivers, \emph{{Single Spin Production Asymmetries from the Hard
  Scattering of Point-Like Constituents}},
  \href{https://doi.org/10.1103/PhysRevD.41.83}{\emph{Phys. Rev.} {\bfseries
  D41} (1990) 83}.

\bibitem{Gutierrez-Reyes:2017glx}
D.~Guti{\'e}rrez-Reyes, I.~Scimemi and A.~A. Vladimirov, \emph{{Twist-2
  matching of transverse momentum dependent distributions}},
  \href{https://doi.org/10.1016/j.physletb.2017.03.031}{\emph{Phys. Lett.}
  {\bfseries B769} (2017) 84}
  [\href{https://arxiv.org/abs/1702.06558}{{\ttfamily 1702.06558}}].

\bibitem{Echevarria:2015uaa}
M.~G. Echevarria, T.~Kasemets, P.~J. Mulders and C.~Pisano, \emph{{QCD
  evolution of (un)polarized gluon TMDPDFs and the Higgs $q_T$-distribution}},
  {\emph{JHEP} {\bfseries 07} (2015) 158}
  [\href{https://arxiv.org/abs/1502.05354}{{\ttfamily 1502.05354}}].

\bibitem{Constantinou:2019vyb}
M.~Constantinou, H.~Panagopoulos and G.~Spanoudes, \emph{{One-loop
  renormalization of staple-shaped operators in continuum and lattice
  regularizations}}, {\emph{Phys. Rev.} {\bfseries D99} (2019) 074508}
  [\href{https://arxiv.org/abs/1901.03862}{{\ttfamily 1901.03862}}].

\bibitem{Green:2020xco}
J.~R. Green, K.~Jansen and F.~Steffens, \emph{{Improvement, generalization, and
  scheme conversion of Wilson-line operators on the lattice in the auxiliary
  field approach}},
  \href{https://doi.org/10.1103/PhysRevD.101.074509}{\emph{Phys. Rev.}
  {\bfseries D101} (2020) 074509}
  [\href{https://arxiv.org/abs/2002.09408}{{\ttfamily 2002.09408}}].

\bibitem{Dotsenko:1979wb}
V.~S. Dotsenko and S.~N. Vergeles, \emph{{Renormalizability of Phase Factors in
  the Nonabelian Gauge Theory}}, {\emph{Nucl. Phys.} {\bfseries B169} (1980)
  527}.

\bibitem{Craigie:1980qs}
N.~S. Craigie and H.~Dorn, \emph{{On the Renormalization and Short Distance
  Properties of Hadronic Operators in {QCD}}}, {\emph{Nucl. Phys.} {\bfseries
  B185} (1981) 204}.

\bibitem{Dorn:1986dt}
H.~Dorn, \emph{{Renormalization of Path Ordered Phase Factors and Related
  Hadron Operators in Gauge Field Theories}}, {\emph{Fortsch. Phys.} {\bfseries
  34} (1986) 11}.

\bibitem{Barone:2001sp}
V.~Barone, A.~Drago and P.~G. Ratcliffe, \emph{{Transverse polarisation of
  quarks in hadrons}},
  \href{https://doi.org/10.1016/S0370-1573(01)00051-5}{\emph{Phys. Rept.}
  {\bfseries 359} (2002) 1}
  [\href{https://arxiv.org/abs/hep-ph/0104283}{{\ttfamily hep-ph/0104283}}].

\bibitem{Luebbert:2016itl}
T.~L{\"u}bbert, J.~Oredsson and M.~Stahlhofen, \emph{{Rapidity renormalized TMD
  soft and beam functions at two loops}}, {\emph{JHEP} {\bfseries 03} (2016)
  168} [\href{https://arxiv.org/abs/1602.01829}{{\ttfamily 1602.01829}}].

\bibitem{Bacchetta:2013pqa}
A.~Bacchetta and A.~Prokudin, \emph{{Evolution of the helicity and transversity
  Transverse-Momentum-Dependent parton distributions}},
  \href{https://doi.org/10.1016/j.nuclphysb.2013.07.013}{\emph{Nucl. Phys. B}
  {\bfseries 875} (2013) 536}
  [\href{https://arxiv.org/abs/1303.2129}{{\ttfamily 1303.2129}}].

\bibitem{Artru:1989zv}
X.~Artru and M.~Mekhfi, \emph{{Transversely Polarized Parton Densities, their
  Evolution and their Measurement}},
  \href{https://doi.org/10.1007/BF01556280}{\emph{Z. Phys. C} {\bfseries 45}
  (1990) 669}.

\bibitem{Gutierrez-Reyes:2018iod}
D.~Gutierrez-Reyes, I.~Scimemi and A.~Vladimirov, \emph{{Transverse momentum
  dependent transversely polarized distributions at
  next-to-next-to-leading-order}},
  \href{https://doi.org/10.1007/JHEP07(2018)172}{\emph{JHEP} {\bfseries 07}
  (2018) 172} [\href{https://arxiv.org/abs/1805.07243}{{\ttfamily
  1805.07243}}].

\bibitem{Chai:2018mwx}
X.~P. Chai, K.~B. Chen and J.~P. Ma, \emph{{A Note on Pretzelosity TMD Parton
  Distribution}},
  \href{https://doi.org/10.1016/j.physletb.2018.12.020}{\emph{Phys. Lett.}
  {\bfseries B789} (2019) 360}
  [\href{https://arxiv.org/abs/1808.10560}{{\ttfamily 1808.10560}}].

\bibitem{Bauer:2000yr}
C.~W. Bauer, S.~Fleming, D.~Pirjol and I.~W. Stewart, \emph{{An Effective field
  theory for collinear and soft gluons: Heavy to light decays}}, {\emph{Phys.
  Rev.} {\bfseries D63} (2001) 114020}
  [\href{https://arxiv.org/abs/hep-ph/0011336}{{\ttfamily hep-ph/0011336}}].

\bibitem{Bauer:2002nz}
C.~W. Bauer, S.~Fleming, D.~Pirjol, I.~Z. Rothstein and I.~W. Stewart,
  \emph{Hard scattering factorization from effective field theory},
  {\emph{Phys. Rev. D} {\bfseries 66} (2002) 014017}
  [\href{https://arxiv.org/abs/hep-ph/0202088}{{\ttfamily hep-ph/0202088}}].

\bibitem{SCETlecture}
C.~W. Bauer and I.~W. Stewart, ``The soft-collinear effective theory.''
  \url{https://courses.edx.org/c4x/MITx/8.EFTx/asset/notes_scetnotes.pdf},
  2014.

\bibitem{Xiong:2013bka}
X.~Xiong, X.~Ji, J.-H. Zhang and Y.~Zhao, \emph{{One-loop matching for parton
  distributions: Nonsinglet case}}, {\emph{Phys. Rev.} {\bfseries D90} (2014)
  014051} [\href{https://arxiv.org/abs/1310.7471}{{\ttfamily 1310.7471}}].

\bibitem{Izubuchi:2018srq}
T.~Izubuchi, X.~Ji, L.~Jin, I.~W. Stewart and Y.~Zhao, \emph{{Factorization
  Theorem Relating Euclidean and Light-Cone Parton Distributions}},
  {\emph{Phys. Rev.} {\bfseries D98} (2018) 056004}
  [\href{https://arxiv.org/abs/1801.03917}{{\ttfamily 1801.03917}}].

\bibitem{Alexandrou:2018eet}
C.~Alexandrou, K.~Cichy, M.~Constantinou, K.~Jansen, A.~Scapellato and
  F.~Steffens, \emph{{Transversity parton distribution functions from lattice
  QCD}}, \href{https://doi.org/10.1103/PhysRevD.98.091503}{\emph{Phys. Rev.}
  {\bfseries D98} (2018) 091503}
  [\href{https://arxiv.org/abs/1807.00232}{{\ttfamily 1807.00232}}].

\bibitem{Liu:2018uuj}
{\scshape Lattice Parton} collaboration, Y.-S. Liu et~al., \emph{{Unpolarized
  isovector quark distribution function from lattice QCD: A systematic analysis
  of renormalization and matching}},
  \href{https://doi.org/10.1103/PhysRevD.101.034020}{\emph{Phys. Rev.}
  {\bfseries D101} (2020) 034020}
  [\href{https://arxiv.org/abs/1807.06566}{{\ttfamily 1807.06566}}].

\bibitem{Liu:2018hxv}
Y.-S. Liu, J.-W. Chen, L.~Jin, R.~Li, H.-W. Lin, Y.-B. Yang et~al.,
  \emph{{Nucleon Transversity Distribution at the Physical Pion Mass from
  Lattice QCD}},  \href{https://arxiv.org/abs/1810.05043}{{\ttfamily
  1810.05043}}.

\bibitem{Korchemsky:1987wg}
G.~P. Korchemsky and A.~V. Radyushkin, \emph{{Renormalization of the Wilson
  Loops Beyond the Leading Order}},
  \href{https://doi.org/10.1016/0550-3213(87)90277-X}{\emph{Nucl. Phys.}
  {\bfseries B283} (1987) 342}.

\end{thebibliography}\endgroup

\end{document}